%% file: Schelling_RandomTags.tex
\documentclass[english,12pt]{article}
\usepackage{array}
\usepackage{pstricks,pst-node,pst-text,pst-3d}
\usepackage{delarray}
\usepackage{t1enc}
\usepackage[utf8]{inputenc}
\usepackage[english]{babel}
\usepackage{graphicx,multirow}
\usepackage{eepic}
\usepackage{epic}
\usepackage{euscript}
\usepackage{indentfirst}
\usepackage{amsmath}
\usepackage{a4wide}
\usepackage{ulem}
\usepackage{url}
\usepackage{amsfonts}
\usepackage{amssymb}
\usepackage{stmaryrd}
\usepackage{theorem}
\usepackage{verbatim}
\usepackage{multirow}
\usepackage[numbers,sort&compress]{natbib}
\usepackage{tabularx}
\usepackage{caption}
\usepackage{subcaption}
\usepackage{float}
\usepackage{pgfplots}

\unitlength=1mm{}

\newcommand{\modNH}[1]{{#1}}

\begin{document}

\title{Random tags can sustain high heterophily levels}

\author{
Julien \textsc{Flaig}\thanks{Corresponding author. Email: \texttt{julien@jflaig.com}.}
\and
Nicolas \textsc{Houy}\thanks{University of Lyon, Lyon, F-69007, France;
CNRS, GATE Lyon Saint-Etienne, F-69130, France. Email: \texttt{houy@gate.cnrs.fr}.}
}

\date{\today}

\maketitle

\begin{abstract}
We consider a spatial model of the emergence of cooperation with synchronous births and deaths. 		
Agents bear a tag and interact with their neighbors by playing the prisoner's dilemma game with strategies depending on their own and opponent's tags.
An agent's payoff determines its chances of reproducing and passing on its strategy.
We show that when tags are assigned at random rather than inherited, a significant heterophilic population of about 40~\% of the whole can emerge and persist. 
Heterophilics defect on agents bearing the same tag as theirs and cooperate with others. 
In our setting, the emergence of heterophily is explained by the correlation between an agent's payoff and its neighbors' payoffs.  
The advantage of heterophily over homophily (cooperating with agents bearing one's tag and defecting with others) when tags are assigned at random makes the emergence of the later an even more interesting phenomenon than previously thought. 
\end{abstract}

\medskip 

Declarations of interest: none.

\pagebreak
\baselineskip=6mm
\section{Introduction}

Why would individuals cooperate at a cost when they can benefit from others' cooperation whatever their own behavior? 
In such situations, individuals have an immediate incentive to free ride on others' cooperation by defecting (not cooperating), although mutual defection may yield lower payoffs than mutual cooperation.
For this reason the emergence and persistence of cooperation in populations has attracted interest from evolutionary biologists and political scientists alike. 

The proverbial prisoner's dilemma game (or close variants) has largely been used to model interactions between two individuals who may either cooperate at a cost, or defect at zero cost.  
We retain the following formulation of the game:
a player who cooperates incurs a cost $c>0$ while his opponent gets a benefit $b > c$. Thus, mutual cooperation yields $b-c>0$ for each player. Free riding is tempting since it yields $b$, but mutual defection yields $0$ for each player. The worse possible outcome is cooperating and being exploited by a free rider: the payoff is then $-c$ for the cooperator. The matrix of the game is: 
\begin{center}
\begin{tabular}{r | c | c |}
	\multicolumn{1}{r}{} & \multicolumn{1}{c}{Cooperate} & \multicolumn{1}{c}{Defect}\\
	\cline{2-3}
	 Cooperate & $b-c$, $b-c$& $-c$, $b$\\
	\cline{2-3}
	 Defect &$b$,$-c$& $0$,$0$\\
	\cline{2-3}
\end{tabular}.
\end{center}

In evolutionary accounts of the emergence of cooperation, an individual's strategy (defect or cooperate) is inherited from his parents, and an individual's reproductive fitness depends on the payoff resulting from interactions with others.   
If individuals interact indiscriminately, exploitation of cooperators by defectors will prevent cooperation to emerge and persist in the population \cite{axelrod1981}.  
A wide range of mechanisms that can uphold cooperation have been proposed, from population viscosity or spatial structure \cite{nowak1992} to repeated interactions \cite{axelrod1981}. We refer to \cite{zaggl2014} and the references therein for a review of these mechanisms.  

One such mechanism is the existence of heritable, recognizable, phenotypic markers, or \textit{tags}, allowing for differentiated strategies.
It was first hypothesized that a gene coding for both a recognizable tag, or \textit{green beard}, and for cooperation with tag-bearing individuals could uphold the emergence of cooperation \cite{hamilton1964a, hamilton1964b, dawkins1976}. 
Instances of such genes in nature, however, remain rare and debated \cite{penn2010}.
The green-beard effect is to be distinguished from the \textit{armpit effect}, where different genes code for the tag (the armpit smell) and for cooperation with tag-bearing individuals \cite{dawkins1982, gardner2010}. 
While the armpit effect can uphold the emergence of cooperation, a population of contingent cooperators would be vulnerable to exploitation by tag-bearing defectors \cite{crozier1986, field2018}. 

Models were subsequently developed to investigate the conditions under which tags can promote the emergence and persistence of cooperation in settings where the prisoner's dilemma game is a model of individual interactions.
In particular, a classic result is that the existence of several tags and spatial structure can lead to the emergence of clusters of like-tagged individuals who cooperate with individuals bearing the same tag and defect against others \cite{hochberg2003, axelrod2004, hammond2006evolution, hammond2006contingent, jansen2006, rousset2007, shultz2009, mcavity2013}. In the following, we refer to such individuals as \textit{homophilic}. 

The emergence of \textit{heterophily}, that is cooperation with individuals bearing a different tag and defection against like-tagged individuals,\footnote{In the literature, heterophilic cooperators are also referred to as \textit{out-group altruists}, \textit{extra-tag altruists}, or \textit{traitorous individuals} \cite{ramazi2018}.} received comparatively less attention than the emergence of homophily.
Indeed, heterophilic cooperation hardly seems sustainable since heterophilic cooperators defect against their own kin bearing the same tag \cite{masuda2015}.
Yet it was shown that heterophily can emerge and persist for large $b/c$ values in a spatially structured population with exactly two tags \cite{laird2011, hadzibeganovic2012}.
In spatial models where individuals are able to choose their location, homophily and heterophily were shown to predominate alternatively as $b/c$ increases \cite{ichinose2015}.
Other authors investigated the role of population structure, modeled by assortment --~an individual's probability of interacting with his own kin \cite{garcia2014}. In their model, heterophilic cooperators are able to emerge and persist by exploiting homophilic cooperators for intermediate assortment.
As for the evolution of homophily and heterophily in well-mixed populations, we refer to \cite{ramazi2018} for a study of the properties of these strategies, and a proof that they cannot persist in the presence of indiscriminate defectors.   
Finally, notice that some simpler evolutionary accounts of homophily and heterophily do not make reference to the prisoner's dilemma game, and disregard interactions between homophilic and heterophilic agents \cite{fu2012, ramazi2016}.

In this article, we bring another example showing that heterophily can indeed emerge under some conditions.
We will see that when tags are not inherited but assigned randomly, heterophily can reach qualitatively important levels, although not over wide parameter ranges. 
We will explain the advantage of heterophily as compared to homophily in terms of payoff correlation.
To our knowledge, payoff correlations have been overlooked in the literature, or at least never been discussed explicitly.
We rely on a standard spatial model with synchronous births and deaths, in which tag-bearing agents interact by playing the prisoner's dilemma game with their neighbors. 
In our base model, we consider the four pure strategies proposed in \cite{axelrod2004}: unconditional defection, heterophily, homophily, and unconditional cooperation.

Our model and the simulation dynamics are introduced  in Section~\ref{sec:model}. 
We present our results in Section~\ref{sec:results}. 
First, we illustrate the emergence of heterophily and perform robustness checks (Section~\ref{sec:emergence}).
Then, we explain the advantage of heterophily over homophily with a model featuring only these two strategies (Section~\ref{sec:explanation}).
Section~\ref{sec:conclu} concludes.

\section{Materials and methods}\label{sec:model}

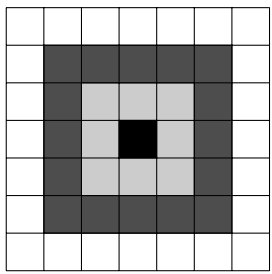
\begin{figure}[ht]
\centering
\input{figs/fig1.tex}
\caption{1-neighborhood of the black cell: the black cell and the light grey cells.
2-neighborhood of the black cell: the black cell, the light grey cells, and the dark grey cells.}\label{fig:1}
\end{figure}

We consider a periodic two-dimensional square lattice with size $n \times n$.\footnote{$n$ will be fixed throughout and, for the sake of simplicity, we will not show it as a parameter everywhere it should be mentioned.} For any cell at coordinates $I=(i,j) \in \{0,\dots,n-1\} \times \{0,\dots,n-1\}$, we call the $r$-neighborhood (with $r \in \mathbb N^{+*}$), denoted $\mathcal N_r(I)$, the Moore neighborhood of $I=(i,j)$ with range $r$.\footnote{%
	Formally, $\mathcal N_k((i,j))$ is defined as the set of cells $(i',j') \in \{0,\dots,n-1\} \times \{0,\dots,n-1\}$ that satisfy:
\begin{itemize}
\item $d(i',i) \leq k$, and 
\item $d(j',j) \leq k$,  
\end{itemize}
where function
$$d(a,a')=\min\left\{\mid(a \mod n)-(a' \mod n)\mid,\mid(a \mod n+n)-(a' \mod n)\mid,\mid(a \mod n-n)-(a' \mod n)\mid\right\}$$
is the periodic distance with periodicity $n$.}
We illustrate neighborhoods in Figure~\ref{fig:1}.

Each cell is occupied by an agent. For the sake of simplicity, we identify cells and agents in our notations. An agent $I$ is endowed with
\begin{itemize}
\item a tag $t_I$ which is an integer between $0$ and $T-1$, where $T$ is the number of possible tags, and
\item a strategy $(s^0_I,s^1_I) \in \{0,1\} \times \{0,1\}$, where $s^0_I$ is the strategy played against agents bearing the same tag, and $s^1_I$ the strategy played against agents bearing a different tag.
\end{itemize}

The simulation dynamics is as follow.

\paragraph{Initialization.} All agents on the grid are given a tag randomly drawn with uniform probability in $\{0, \dots ,T-1\}$. 
Unless otherwise specified, they are also given a random strategy drawn with uniform probability in $\{0,1\} \times \{0,1\}$.

\paragraph{Interactions.} In each generation, all agents interact with all other agents in their $r$-neighborhood excluding themselves.
Let agent $I$ with tag $t_I$ and strategy $(s^0_I,s^1_I)$ interact with agent $J$ with tag $t_J$ and strategy $(s^0_J,s^1_J)$. 
From this interaction, agent $I$ earns payoff
\[
	\left\{\begin{array}{ll}
		-c.s^0_I+b.s^0_J & \text{if~} t_I=t_J \\
		-c.s^1_I+b.s^1_J & \text{otherwise.}
	\end{array}\right.
\]
The total payoff obtained by agent $I$ through interactions with its neighbors is denoted by $\pi_I$. 

\paragraph{Reproduction and inheritance.} 
Once all agents have interacted with their neighbors, a new population is generated. The total payoff $\pi_I$ of any agent $I$ determines its reproductive fitness $f_I = \exp(\beta.\pi_I)$. 
The inverse temperature parameter $\beta$ represents the strength of selection pressure.
For each cell $(k,l)$, a parent $I$ is randomly picked in the $r$-neighborhood $\mathcal N_r((k,l))$ with probability proportional to its reproductive fitness $f_I$.
Hence, the probability of being picked for reproduction among the agents of a given $r$-neighborhood is given by a softmax function of the total payoffs in the neighborhood \cite{szabo1998}. 
In the limit $\beta \rightarrow \infty$, the agents with the highest payoff are selected for reproduction. For $\beta=0$, payoffs are irrelevant and parents are picked with equal probability.
We then consider two treatments:
\begin{itemize}
\item \textit{Inherited tag treatment.} \modNH{The new agent inherits its parent's tag. With mutation probability $\mu_T$, mutants are given a random tag drawn with uniform probability.}
\item \textit{Random tag treatment.} \modNH{The new agent is given a random tag drawn with uniform probability.}
\end{itemize}
\modNH{In all treatments, the new agent inherits its parent's strategy with mutation probability $\mu_S$. Mutants are given a random strategy drawn with uniform probability in $\{0,1\} \times \{0,1\}$.
In the following, we will consider an equal probability of mutation for tags and strategies in the inherited tag treatment and denote $\mu_S=\mu_T=\mu$.
In the random tag treatment, $\mu_S=\mu$ denotes the strategy mutation probability.
In the inherited tag treatment, tag and strategy mutations are independent.}

\modNH{Unless stated otherwise, simulations are run for $G$ generations which we checked is large enough for stationary states to be reached.} In Table~\ref{tab:1}, we display the default parameter values for our simulations.

\begin{table}[ht]
\centering
\input{figs/tab1.tex}
\caption{Parameters and default value.}\label{tab:1}
\end{table}

\section{Results}\label{sec:results}

\subsection{Emergence of heterophily}
\label{sec:emergence}

\begin{figure}[ht]
\centering
\begin{subfigure}{.45\textwidth}
  \centering
  \includegraphics[width=.9\linewidth]{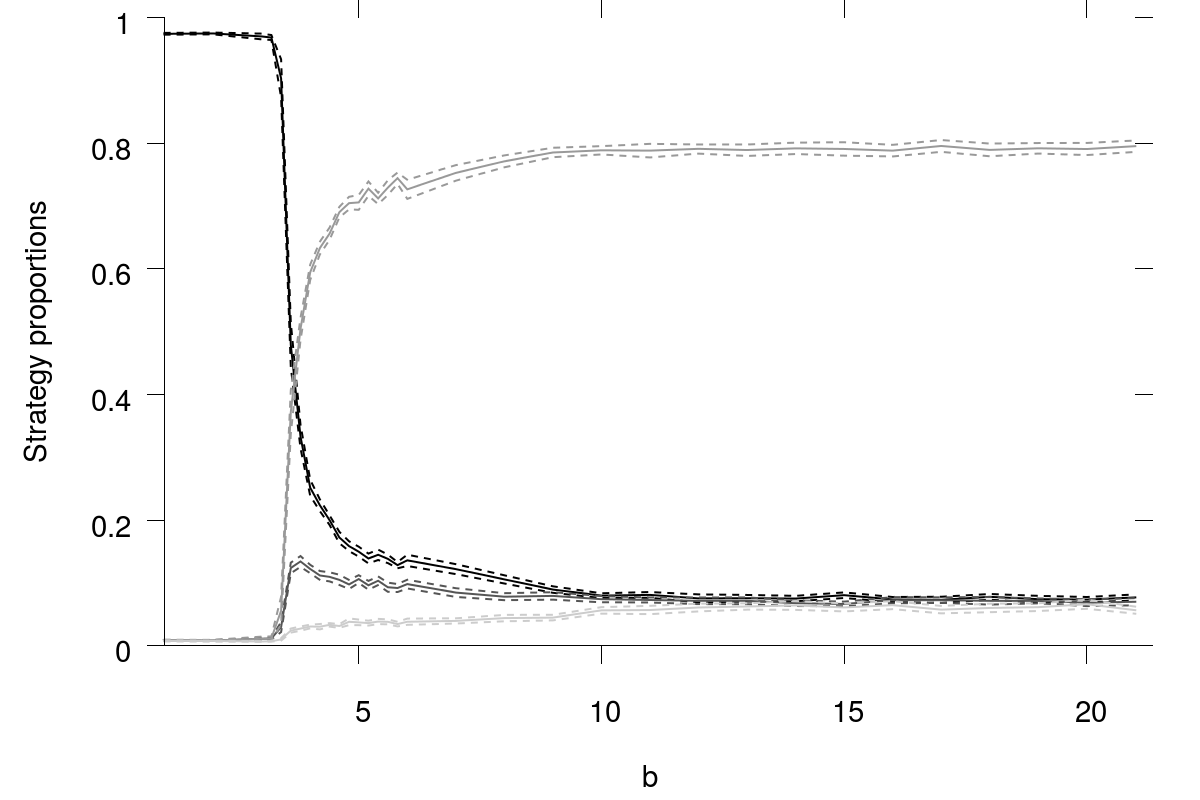}
  \caption{Inherited tags}\label{fig:bvarin}
\end{subfigure}%
\begin{subfigure}{.45\textwidth}
  \centering
  \includegraphics[width=.9\linewidth]{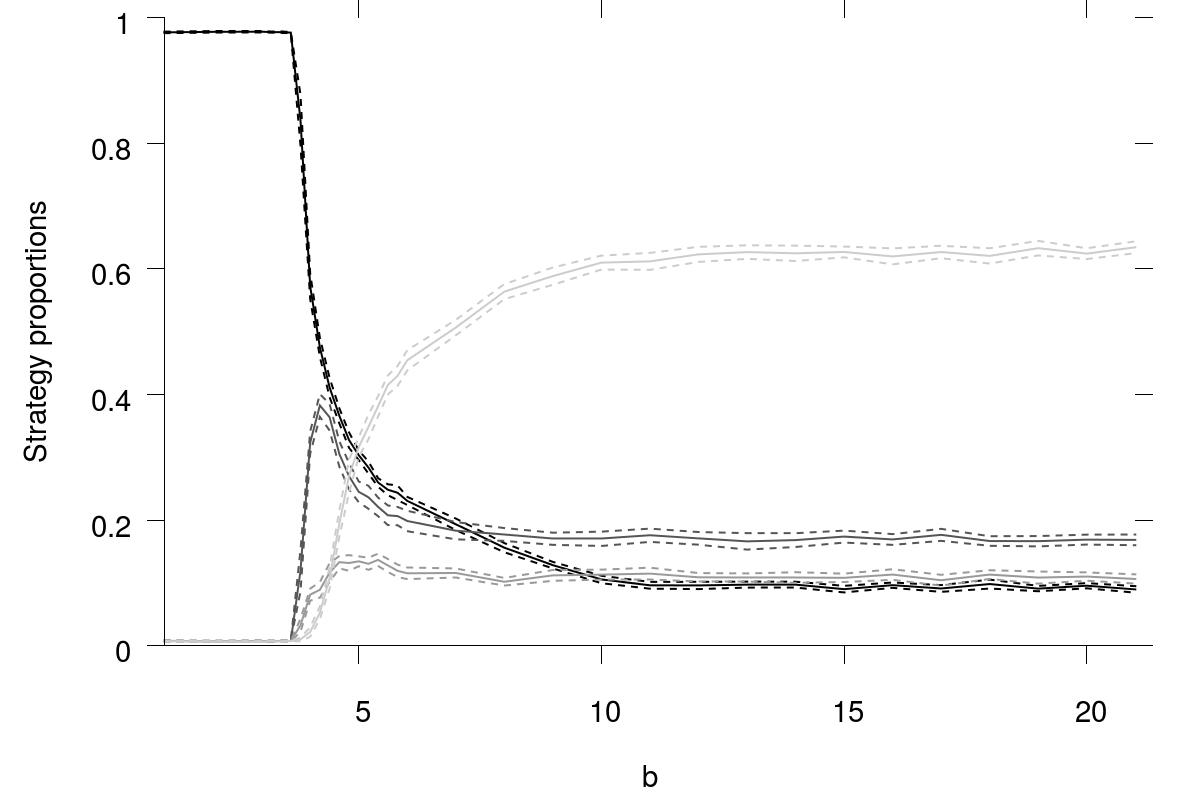}
  \caption{Random tags}\label{fig:bvarra}
\end{subfigure}%
\caption{Evolution of the proportion of each strategy at generation 20,000 depending on $b$ and for the different treatments. From darkest to lightest grey: strategies $(0,0)$, $(0,1)$, $(1,0)$, $(1,1)$. Plain lines are averages and dashed lines are 95~\% confidence intervals for 24 simulation runs.}\label{fig:bvar}
\end{figure}

Figure \ref{fig:bvar} shows the proportion of each strategy in the population at generation 20,000 as a function of the benefit of cooperation $b$, with other parameters set to their default value (Table~\ref{tab:1}).
Our results in the inherited tag treatment (Figure~\ref{fig:bvarin}) are consistent with the classic results of the tag-based cooperation literature \cite{axelrod2004, laird2011}. 
For low values of $b$, unconditional defection (strategy $(0,0)$) predominates.
For high values of $b$, homophily (strategy $(1,0)$) predominates as homophilic agents coordinate with their own kin and form patches of like-tagged cooperating agents.  
This, obviously, cannot hold when tags are not inherited but assigned randomly (Figure~\ref{fig:bvarra}). In this case, unconditional cooperation predominates for large values of $b$, while unconditional defection still predominates for low values of $b$.   
Interestingly, relatively high levels of heterophily (strategy $(0,1)$) are reached for $b$ around $4.2$ when tags are assigned at random. 

\begin{figure}[ht]
\centering
\begin{subfigure}{.45\textwidth}
  \centering
  \includegraphics[width=.9\linewidth]{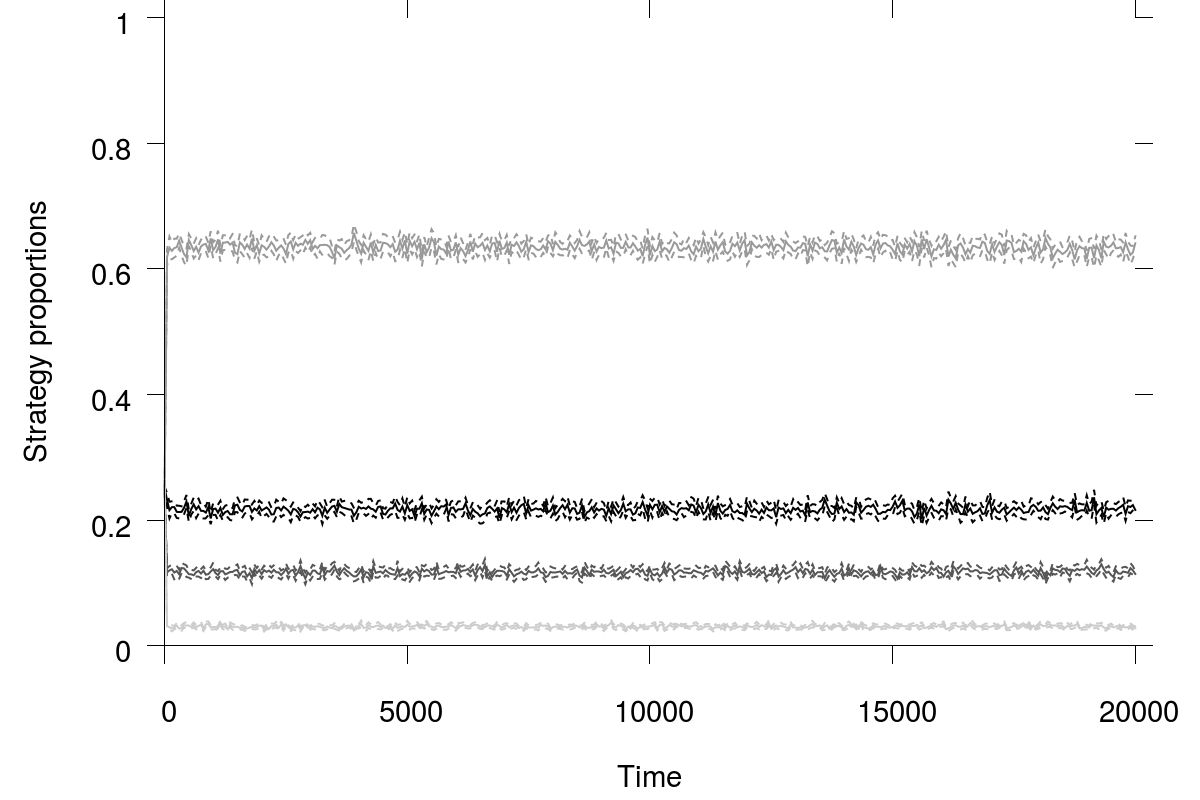}
  \caption{Inherited tags}\label{fig:genin}
\end{subfigure}%
\begin{subfigure}{.45\textwidth}
  \centering
  \includegraphics[width=.9\linewidth]{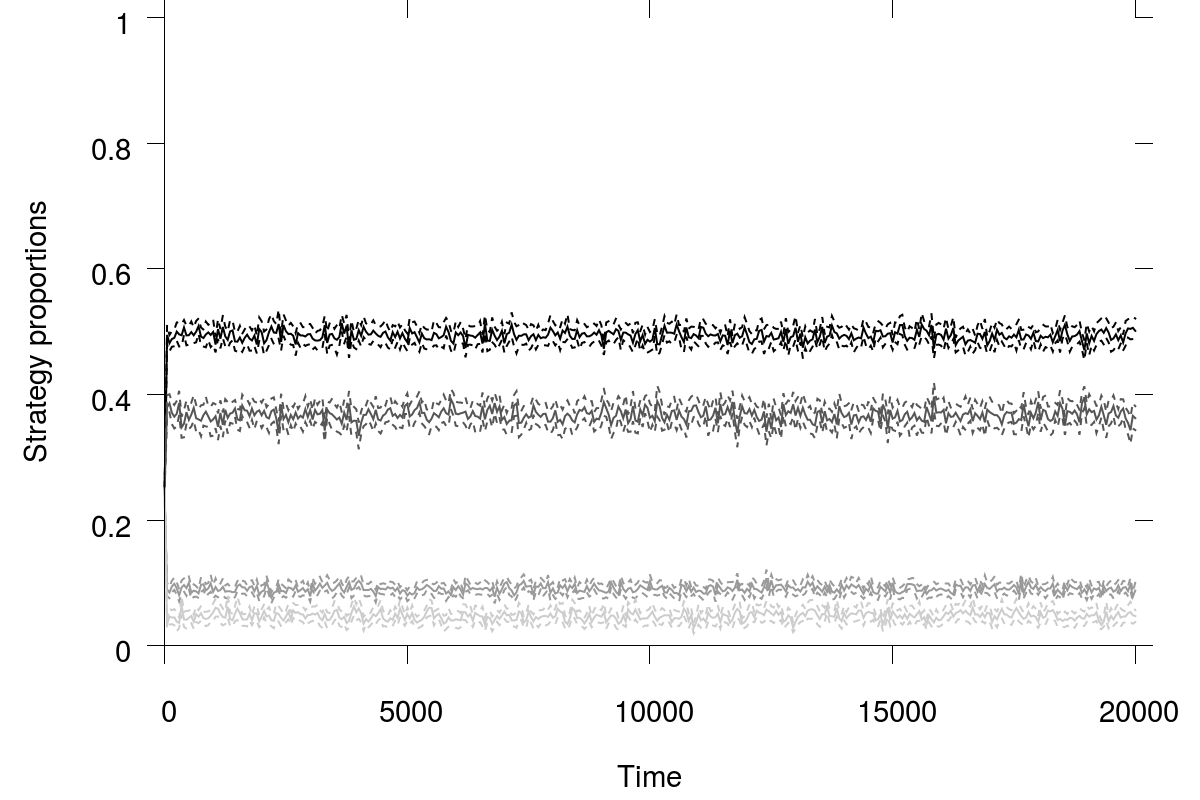}
  \caption{Random tags}\label{fig:genra}
\end{subfigure}%
\caption{Evolution of the proportion of each strategy for simulations with the base case parameters and for the different treatments. From darkest to lightest grey: strategies $(0,0)$, $(0,1)$, $(1,0)$, $(1,1)$. Plain lines are averages and dashed lines are 95~\% confidence intervals for 24 simulation runs.}\label{fig:gen}
\end{figure}

This is best illustrated in Figure~\ref{fig:gen} where we show the evolution of the proportion of each strategy over generations for $b=4.2$. 
Stationary states are obtained very fast.
When tags are assigned randomly (Figure~\ref{fig:genra}), heterophilic agents represent close to 40~\% of the population, just below unconditional defectors (about 50~\% of the population). 
This is to be contrasted with the inherited tag treatment (Figure~\ref{fig:genin}), in which heterophily is almost not present (about 10~\% of the population).
In the following, $b$ is set to default value 4.2.

The emergence of heterophily in the random tag treatment still holds when all agents in the initial generation are unconditional defectors (see Figure~\ref{fig:2R1} in Appendix). 
The result also holds for neighborhood size $r=2$ (see Figure~\ref{fig:neighbSize} in Appendix).
Yet more thorough robustness checks show that heterophily only emerges in relatively small ranges of parameter values. 
This has already been seen for parameter $b$ when other parameters are set to their default value (Figure~\ref{fig:bvarra}).
Similarly, our result only holds for a restricted range of inverse temperatures $\beta$ and of strategy mutation probabilities $\mu$ when other parameters are set to their default value (Figures~\ref{fig:tempvar} and~\ref{fig:muvar} in Appendix).
In the $\beta b$-plane, a maximal heterophilic population around 30--40~\% of the whole is reached along a crest (Figure~\ref{fig:btempra01} in Appendix).
The range of $b$ values for which high levels of heterophily are reached is smaller for the small values of $\mu$ typically found in the literature \cite{axelrod2004, laird2011}.   
Finally, the result does not hold in a model implementing Von Neumann neighborhoods (Figure~\ref{fig:cross}).

Still, the advantage of heterophily in the random tag treatment remains non negligible in magnitude, and as we will now illustrate, its cause is worth being explained further.

\clearpage
\subsection{Explanation}
\label{sec:explanation}

\begin{figure}[ht]
\centering
\begin{subfigure}{.45\textwidth}
  \centering
  \includegraphics[width=.9\linewidth]{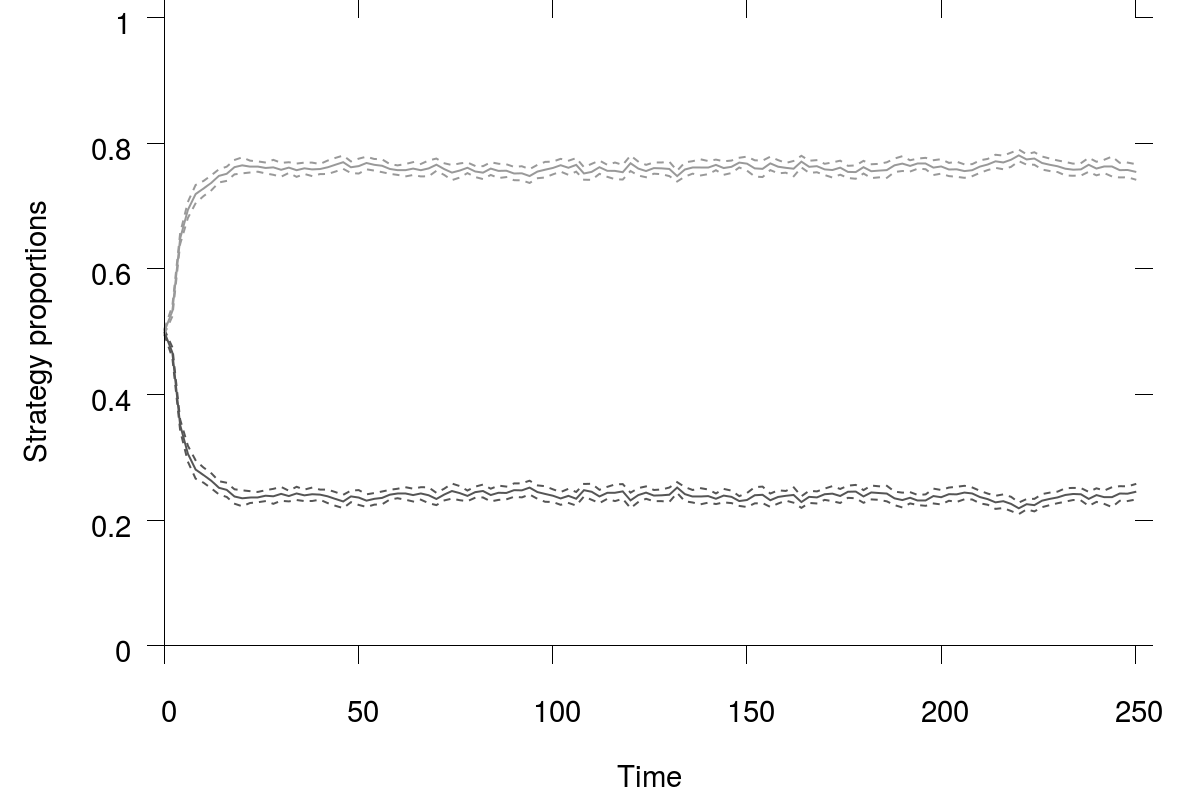}
  \caption{Inherited tags}\label{fig:7i2}
\end{subfigure}%
\begin{subfigure}{.45\textwidth}
  \centering
  \includegraphics[width=.9\linewidth]{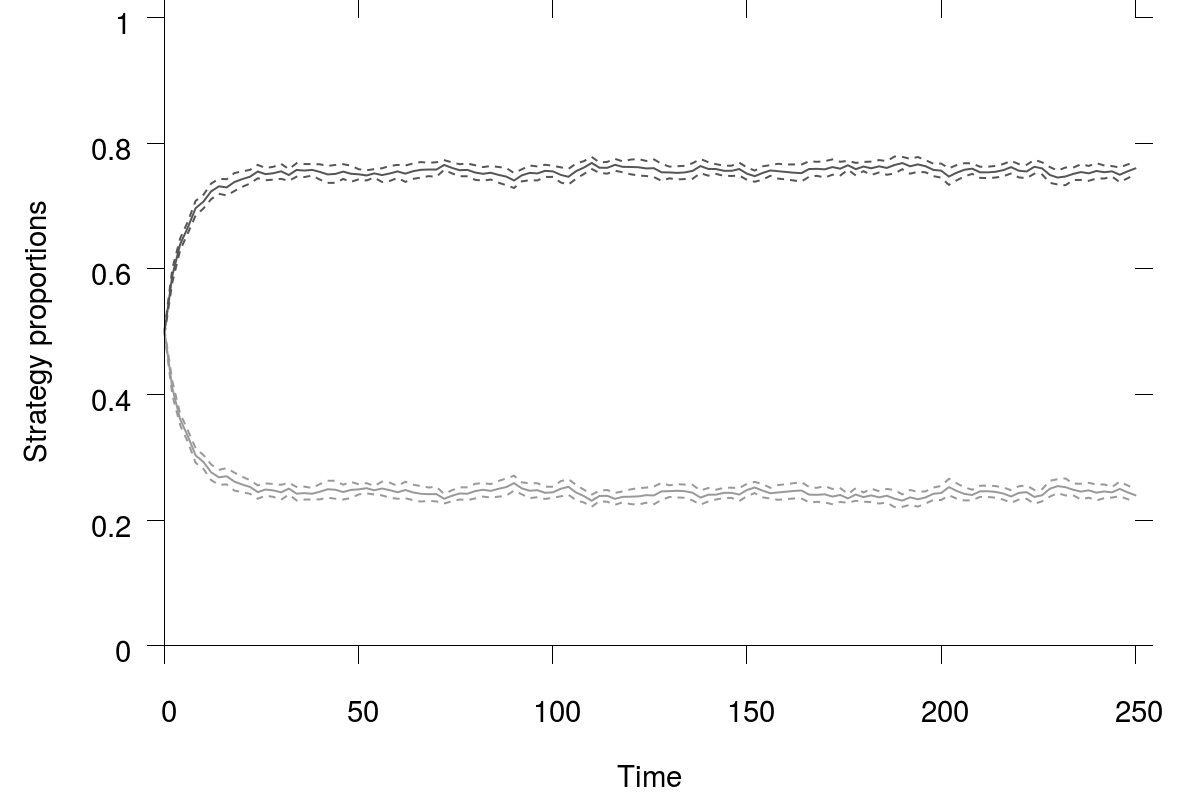}
  \caption{Random tags}\label{fig:7i1}
\end{subfigure}%

\caption{Evolution of the proportion of each strategy for simulations with the base case parameters, for the different treatments and with strategies $(0,1)$ (dark grey) and $(1,0)$ (light grey) only. Plain lines are averages and dashed lines are 95~\% confidence intervals for 24 simulation runs.}\label{fig:2strats}
\end{figure}

In order to explain the evolutionary advantage of heterophily, let us simplify a bit our framework and allow only for heterophily and homophily.
In Figure~\ref{fig:2strats}, we show the results obtained with the same model and parameter values as previously, but restricting the set of strategies to $\{(0,1), (1,0)\}$.
In the inherited tag treatment (Figure~\ref{fig:7i2}), homophilic agents are still able to form clusters of like-tagged cooperators and come to predominate.
In the random tag treatment (Figure~\ref{fig:7i1}), heterophilic agents predominate. We also find that heterophily can emerge and persist for any value of $b$, which suggests that the emergence of heterophily is only prevented for values of $b$ outside a small range by the presence of unconditional cooperators and defectors in our base model (see Figure~\ref{fig:bvarra}).
Importantly, in the the random tag treatment, heterophily predominates from the very first generation.  
This contrasts with the emergence of homophily in the inherited tag treatment, and with the emergence of heterophily reported in previous studies \cite{laird2011}, which unfold over several generations.

That heterophilic agents perform comparatively better in the random tag treatment than in the inherited tag treatment, while homophilic agents perform comparatively worse, comes as no surprise.
In the initial state in both treatments, the expected payoff per interaction of any agent is $(b-c)/2$.
In the inherited tag treatment, heterophilic agents get a payoff of $0$ when interacting with their kin, while homophilic agents get a payoff of $b-c$ when interacting with their own and are able to form clusters. 
By contrast, in the random tag treatment, tags bring no information about kinship (all agents have a $1/2$ probability of bearing either tag), thus both heterophilic and homophilic agents have an expected payoff of $(b-c)/2$ when interacting with their kin. 
But following this reasoning, one would expect equal levels of homophily and heterophily in the random tag treatment, which does not match observation where heterophily is at an advantage even in the initial state (Figure~\ref{fig:7i1}).

Reasoning only in terms of expected payoffs actually misses one important point, that is correlations between payoffs.
In fact, the number of offspring of an agent does not only depend on its own payoff, but also on the payoff of its neighbors with which it will compete.
To see this, let us focus on the initial state in the random tag treatment.
As noted above, the expected payoff per interaction for both heterophilic and homophilic agents is $(b-c)/2 = 1.6$, that is a total expected payoff of $12.8$ for 8 interactions.
Similarly, the expected total payoff of an heterophilic's neighbor is 12.8 just as that of an homophilic's neighbor.
We generated 1,000 initial states and found an average payoff of 12.802 with 95~\% confidence interval (CI) 12.791--12.812 for heterophilic agents, and of 12.792 (95~\% CI: 12.781--12.803) for homophilic agents.
The mean observed payoff of heterophilic agents' neighbors is 12.795 (95~\% CI: 12.792--12.798), and the mean observed payoff of homophilic agents' neighbors is 12.798 (95~\% CI: 12.795--12.801).
Hence, not only have homophilic and heterophilic agents the same expected payoff, but their neighbors also have the same expected payoffs.
This fact, however, can be misleading as it tends to hide the correlations between and agent's payoff and its neighbors'.

\begin{figure}[ht]
\centering
  \includegraphics[width=.9\linewidth]{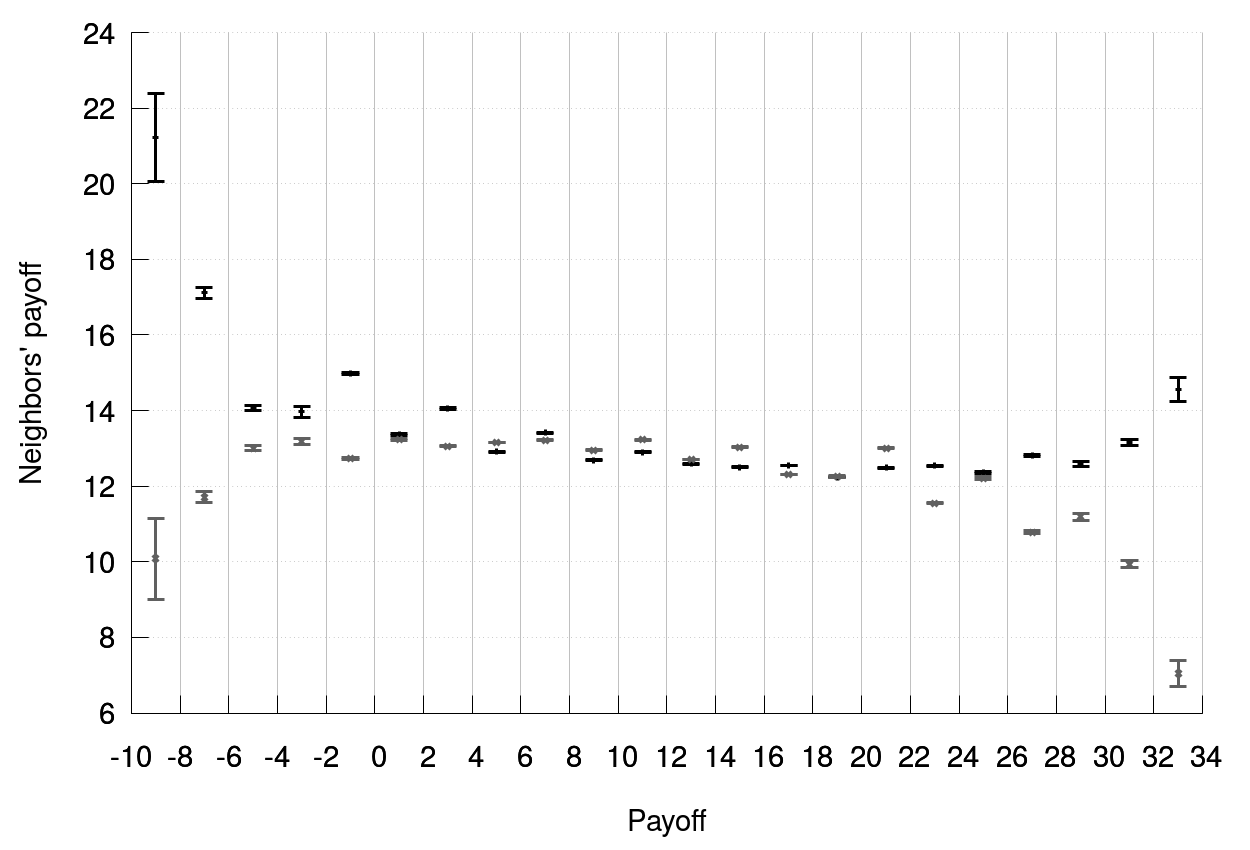}
\caption{Average payoffs of an agent's neighbors as a function of the payoff of the agent itself, depending on it having strategy $(0,1)$ (black) or $(1,0)$ (grey). Whiskers: 95~\% CI with 1,000 simulations of the initial state. Vertical grey lines separate the payoff bins.}\label{fig:correl}
\end{figure}

In Figure~\ref{fig:correl}, we show the average payoffs of an agent's neighbors as a function of the payoff of the agent itself, computed over the 1,000 initial states.\footnote{Strictly speaking, agents also compete with the neighbors of their direct neighbors. Unsurprisingly, the correlation between an agent payoff and the payoffs of agents at a periodic distance of 2 is small. See Figure~\ref{fig:correl12} in Appendix.}
We see that, first, an agent's payoff is indeed correlated to the average payoff of its neighbors, and, second, that this correlation is different for heterophilic and homophilic agents.  
The differences in payoff correlations translate into differences in offspring.
Here, the observed mean number of offspring is 1.121 (95~\% CI: 1.118--1.125) for heterophilic agents and only 0.879 (95~\% CI: 0.875--0.882) for homophilic agents.
In the random tag treatment, this advantage of heterophily remains in the subsequent generations, giving rise to the preponderance of heterophily.

\section{Conclusion}\label{sec:conclu}

A rich literature has shown how the introduction of inheritable tags can promote the emergence of homophilic cooperation when individual interactions are modeled by the prisoner's dilemma game. 
In comparison to homophily, heterophily has received very little attention. 

In this article, we brought an additional example of emergence of heterophilic cooperation.
We used a standard spatial model in which agents bear tags and interact with their neighbors by playing the prisoner's dilemma, using inherited strategies that are contingent on their own and opponent's tags.  
We showed that heterophily can emerge when tags are assigned randomly to agents rather than inherited.
Although the emergence of heterophily does not occur over wide ranges of parameter values and highly depends on the reproduction process, our result has some far-reaching implications.

First, the emergence of heterophily is qualitatively non negligible. In our instance, heterophilic agents can make up about 40~\% of the whole population.

Second, the emergence of heterophily is explained by the correlation between an agent's payoff and his neighbors' payoffs, and by the corresponding differences in fitness between neighbors. 
Such correlations are usually overlooked in the literature. 
Here, it is necessary to look beyond averages, even conditional on strategies, to explain the result.
We believe that our results and the underlying mechanisms could prevail more generally and have a broader significance.

Finally, our work suggests a change in perspective.  
In the tag-based cooperation literature, the emergence of homophilic cooperation in the presence of inheritable tags is often compared to its non-emergence in the absence of tags. 
But, as we have shown, payoff correlations give rise to an advantage of heterophily over homophily when tags are assigned at random.
If we accept random tags and heterophily (another form of cooperation!) as reference point, then the emergence of homophilic cooperation seems all the more puzzling and a relevant subject of inquiry.

\pagebreak

\bibliography{Schelling}

\pagebreak

\clearpage
\appendix
\setcounter{figure}{0}
\renewcommand\thefigure{App-\arabic{figure}}    
\setcounter{table}{0}
\renewcommand\thetable{App-\arabic{table}}    
\setcounter{page}{1}

\part*{Appendix}

\section{Additional results}

\subsection{Change in initial state}

\begin{figure}[ht]
\centering
\begin{subfigure}{.45\textwidth}
  \centering
  \includegraphics[width=.9\linewidth]{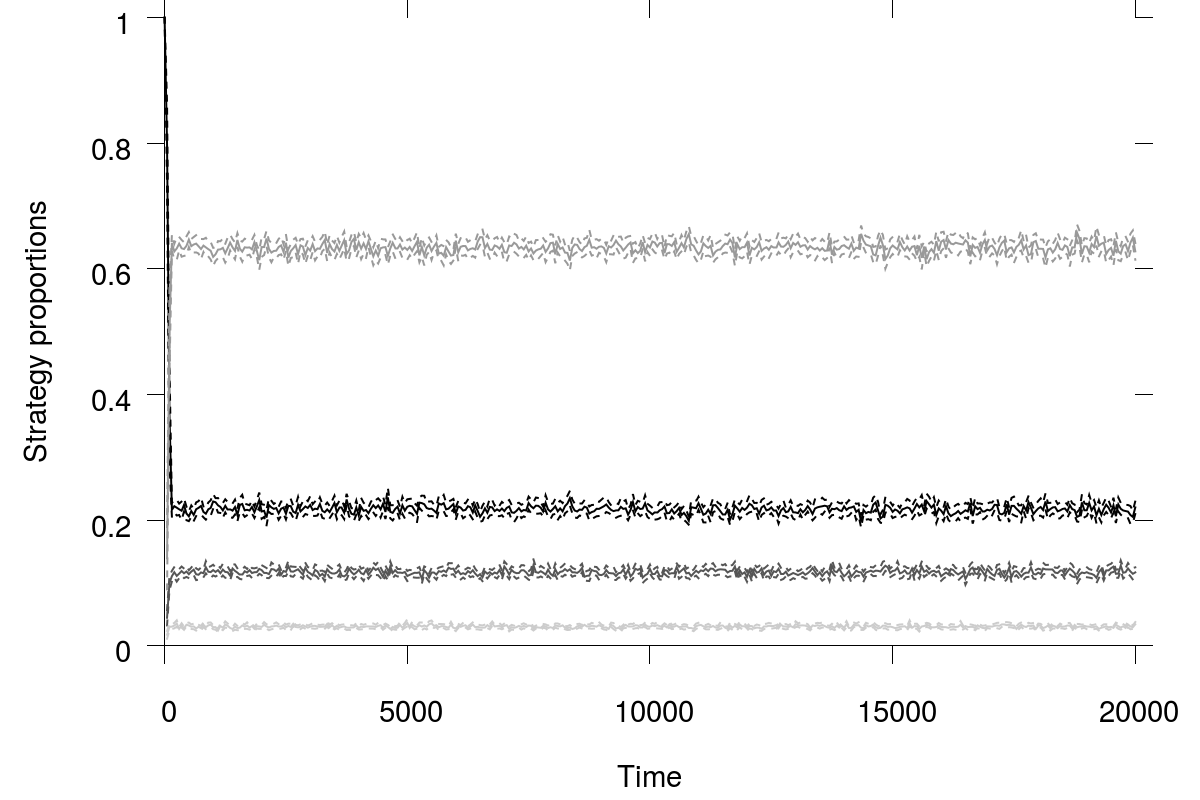}
  \caption{Inherited tags}\label{fig:2R1i2}
\end{subfigure}%
\begin{subfigure}{.45\textwidth}
  \centering
  \includegraphics[width=.9\linewidth]{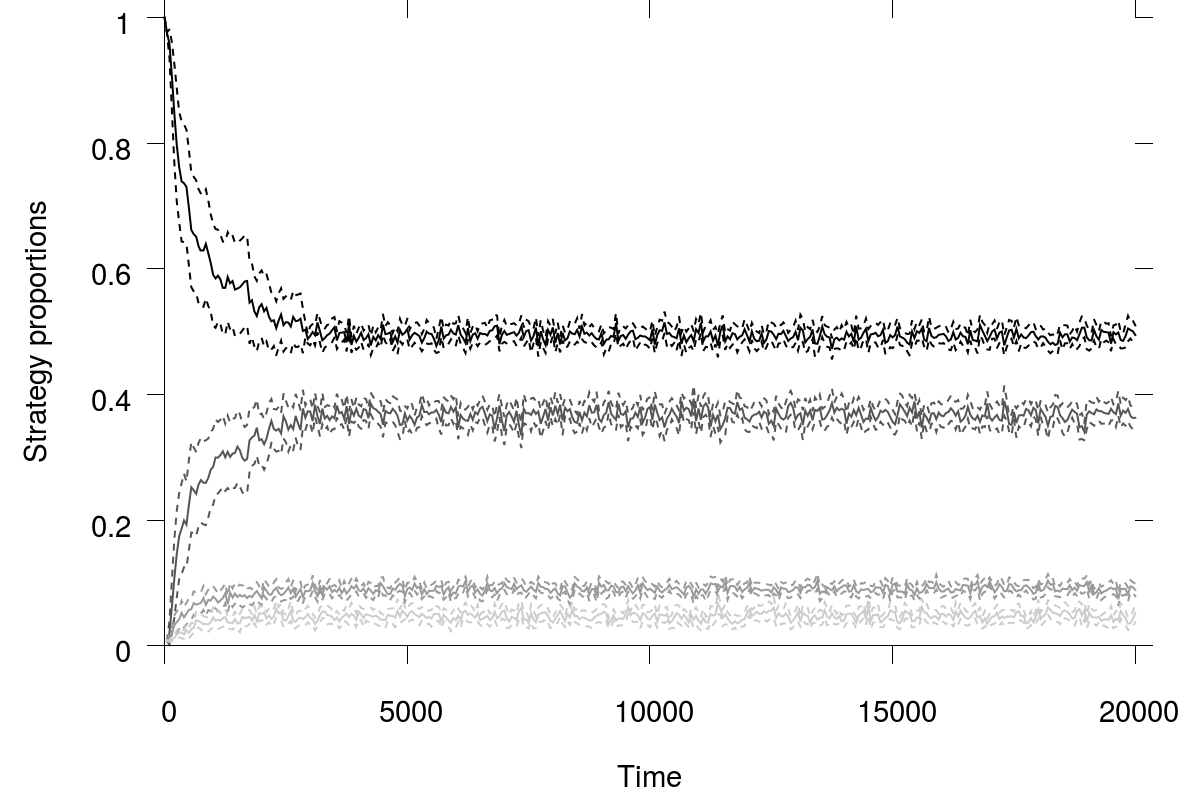}
  \caption{Random tags}\label{fig:2R1i1}
\end{subfigure}%
\caption{Same as Figure \ref{fig:gen} with initial strategy $(0,0)$ for all individuals.}\label{fig:2R1}
\end{figure}

\subsection{Change in neighborhood definition}

\begin{figure}[ht]
\centering
\begin{subfigure}{.45\textwidth}
  \centering
  \includegraphics[width=.9\linewidth]{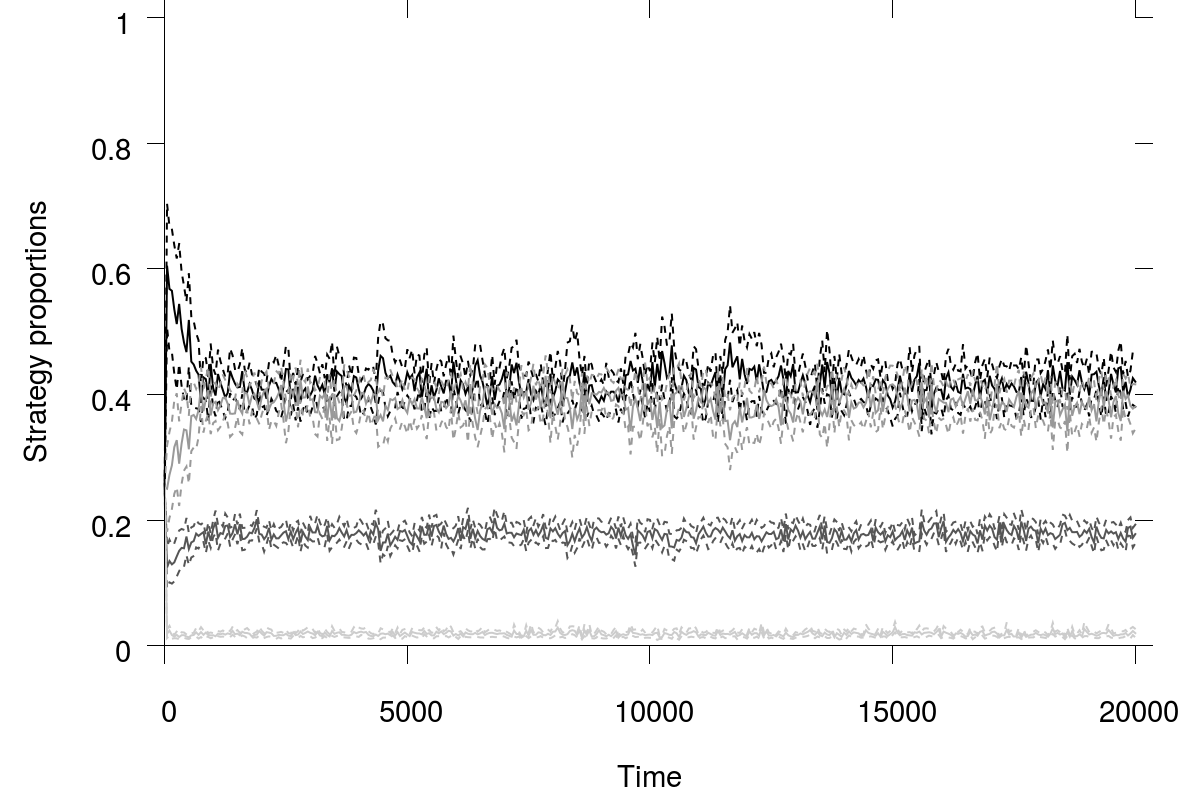}
  \caption{Inherited tags}\label{fig:neighbSize2}
\end{subfigure}%
\begin{subfigure}{.45\textwidth}
  \centering
  \includegraphics[width=.9\linewidth]{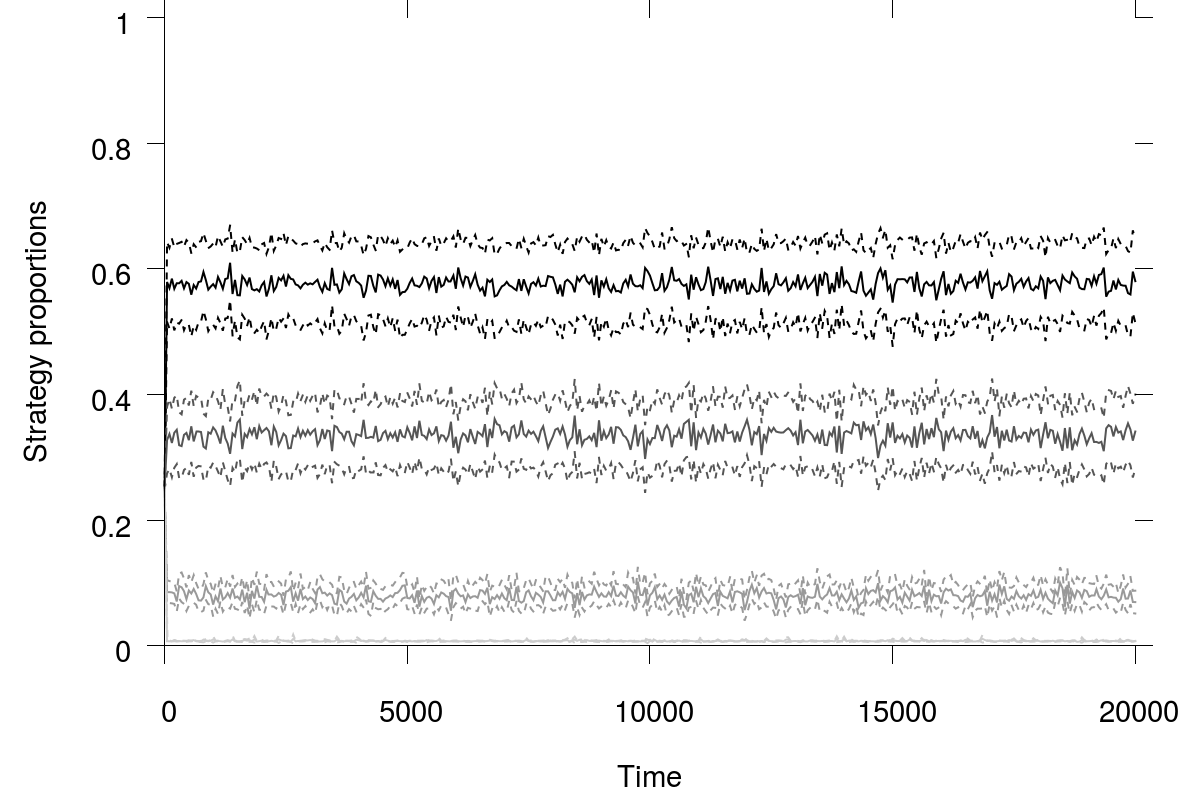}
  \caption{Random tags}\label{fig:neighbSize1}
\end{subfigure}%
\caption{Same as Figure \ref{fig:gen} with neighborhoods of size 2.}\label{fig:neighbSize}
\end{figure}

\begin{figure}[ht]
\centering
\begin{subfigure}{.45\textwidth}
  \centering
  \includegraphics[width=.9\linewidth]{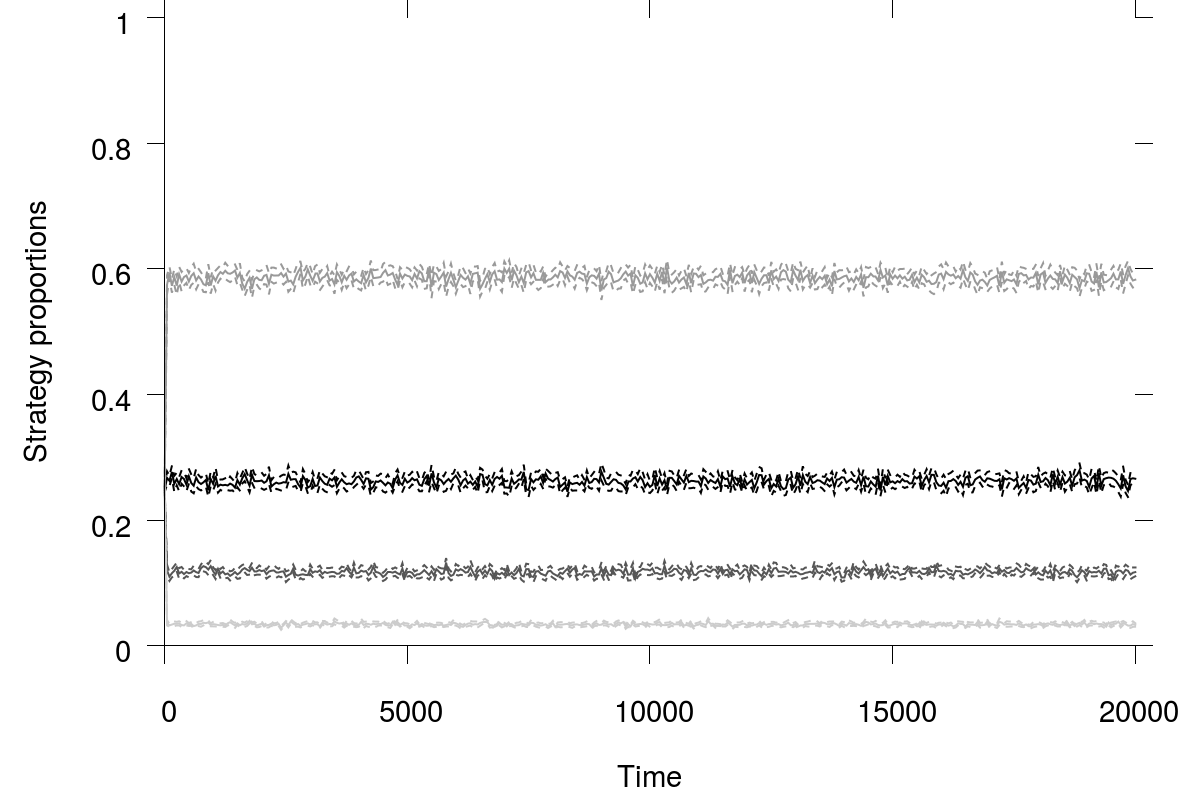}
  \caption{Inherited tags}\label{fig:cross2}
\end{subfigure}%
\begin{subfigure}{.45\textwidth}
  \centering
  \includegraphics[width=.9\linewidth]{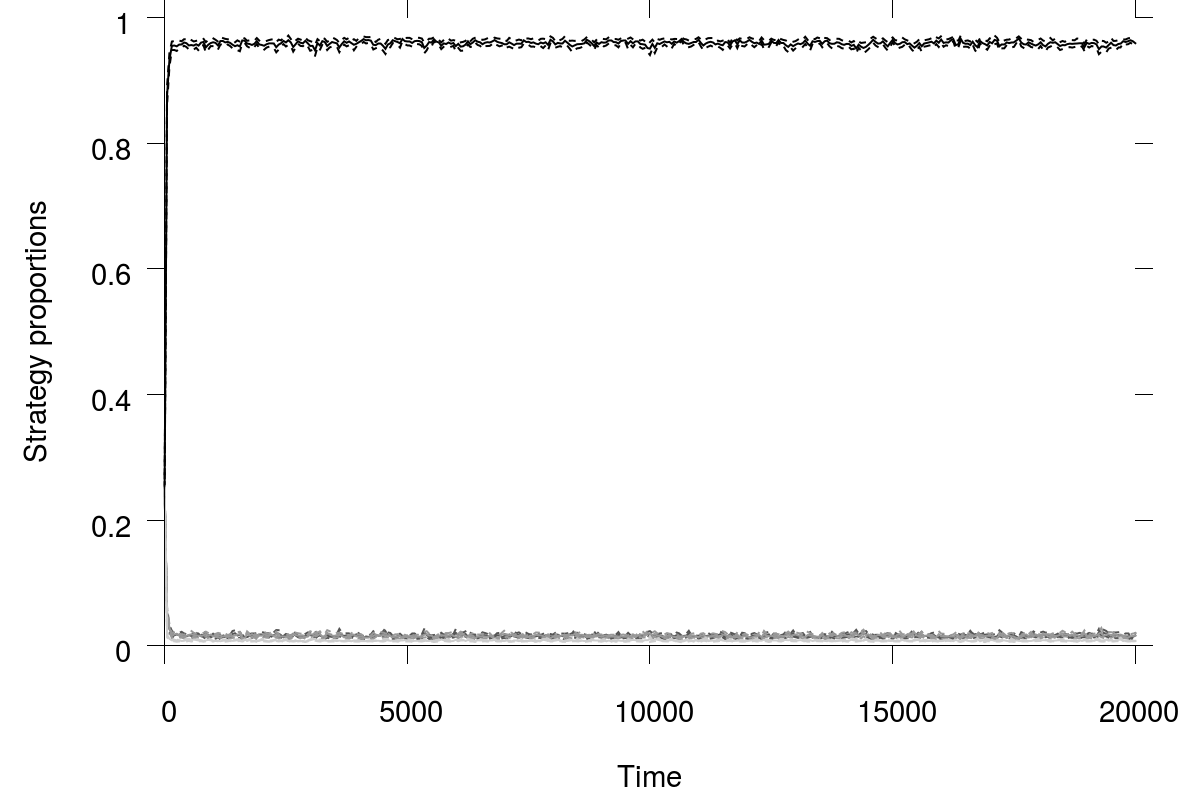}
  \caption{Random tags}\label{fig:cross1}
\end{subfigure}%
\caption{Same as Figure \ref{fig:gen} with obtained with Von Neumann neighborhoods (North-South-East-West adjacent cells).}\label{fig:cross}
\end{figure}

From Figure \ref{fig:cross}: our result for random tags is not robust for another definition of neighborhood (but we only check one dimensional changes). Nothing changes for inherited tags (Figure \ref{fig:cross2}).

\clearpage
\subsection{Change in inverse temperature}

%
%

\begin{figure}[ht]
\centering
\begin{subfigure}{.45\textwidth}
  \centering
  \includegraphics[width=.9\linewidth]{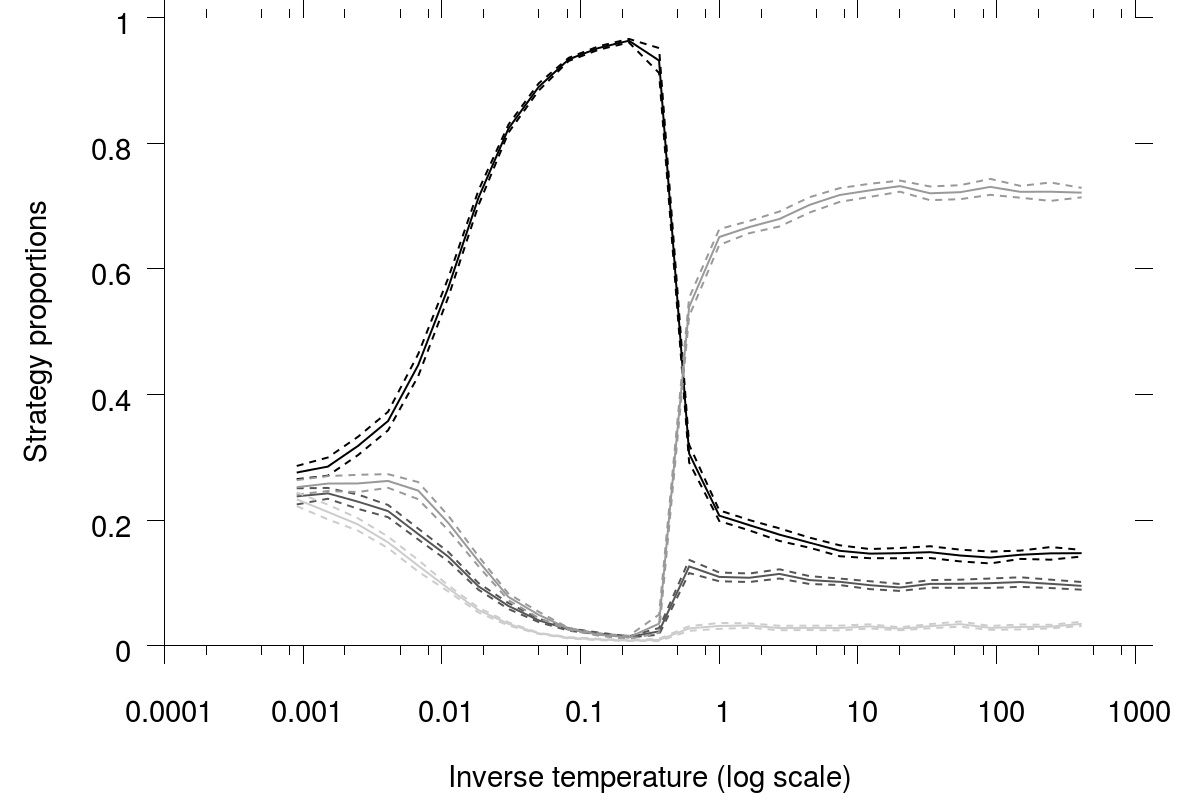}
  \caption{Inherited tags}\label{fig:3i2}
\end{subfigure}%
\begin{subfigure}{.45\textwidth}
  \centering
  \includegraphics[width=.9\linewidth]{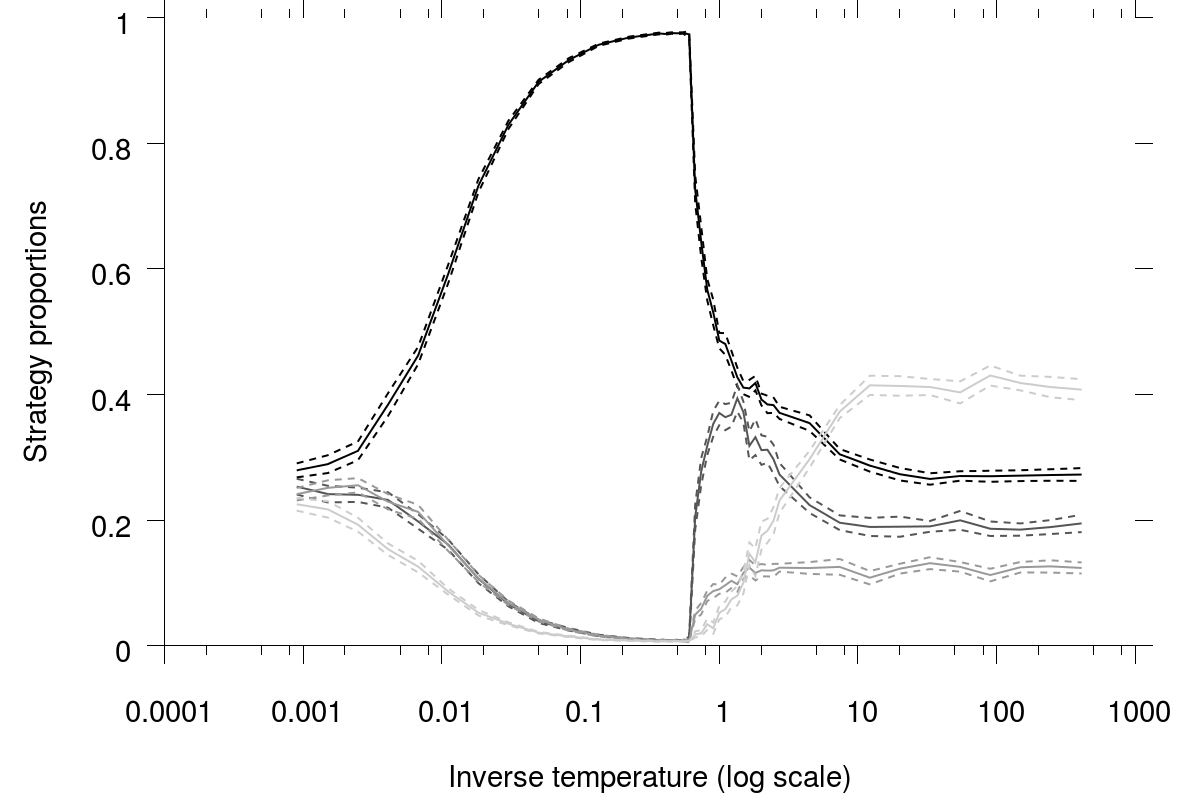}
  \caption{Random tags}\label{fig:3i1}
\end{subfigure}%
\caption{Evolution of the proportion of each strategy at generation 20,000 depending on $\beta$ and for the different treatments. From darkest to lightest grey: strategies $(0,0)$, $(0,1)$, $(1,0)$, $(1,1)$. Plain lines are averages and dashed lines are 95~\% confidence intervals for 24 simulation runs.}\label{fig:tempvar}
\end{figure}

\begin{figure}[ht]
\centering
\begin{subfigure}{.45\textwidth}
  \centering
  \includegraphics[width=.9\linewidth]{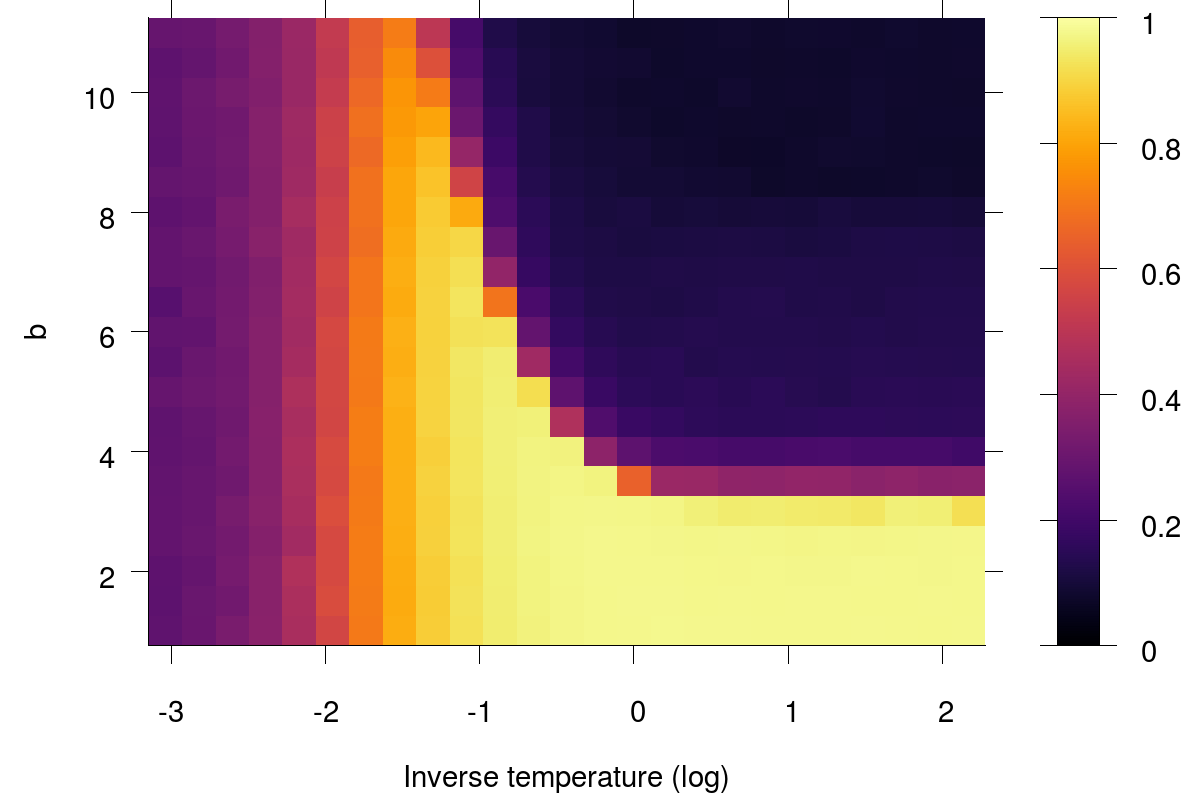}
  \caption{Inherited tags, strategy $(0,0)$}\label{fig:btempin00}
\end{subfigure}%
\begin{subfigure}{.45\textwidth}
  \centering
  \includegraphics[width=.9\linewidth]{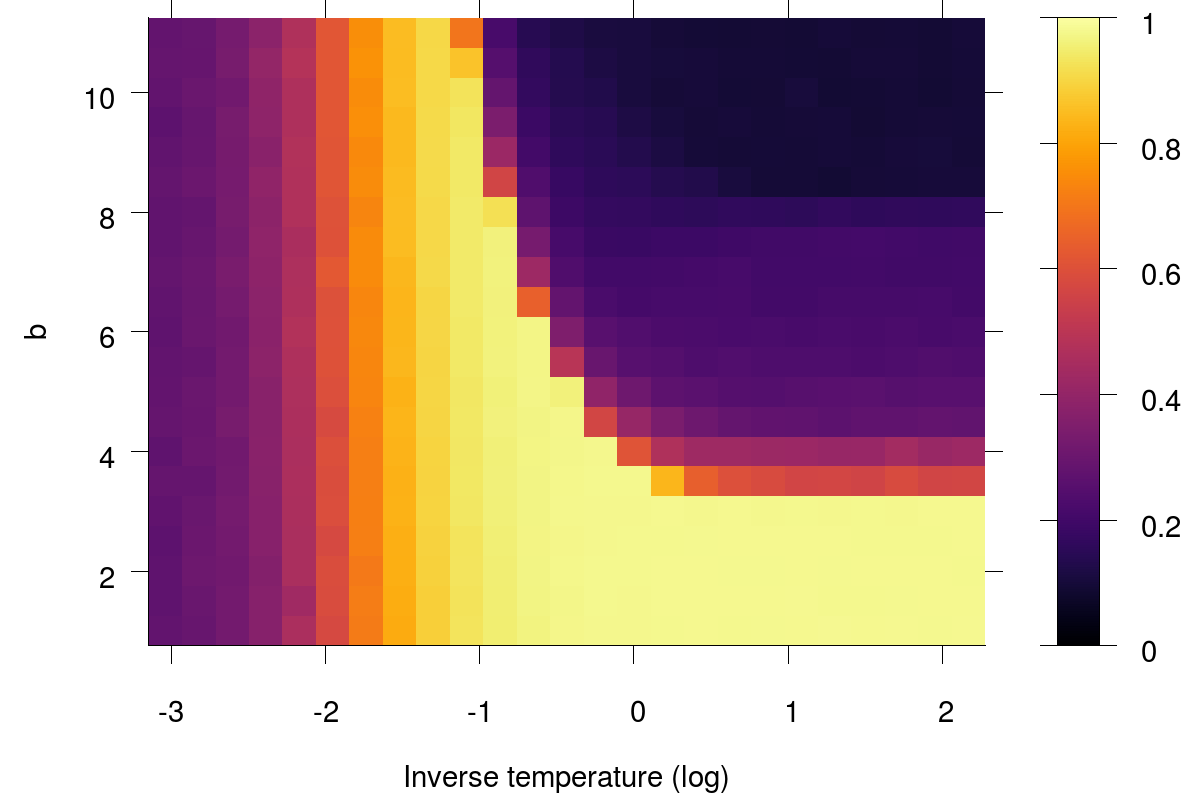}
  \caption{Random tags, strategy $(0,0)$}\label{fig:btempra00}
\end{subfigure}%

\begin{subfigure}{.45\textwidth}
  \centering
  \includegraphics[width=.9\linewidth]{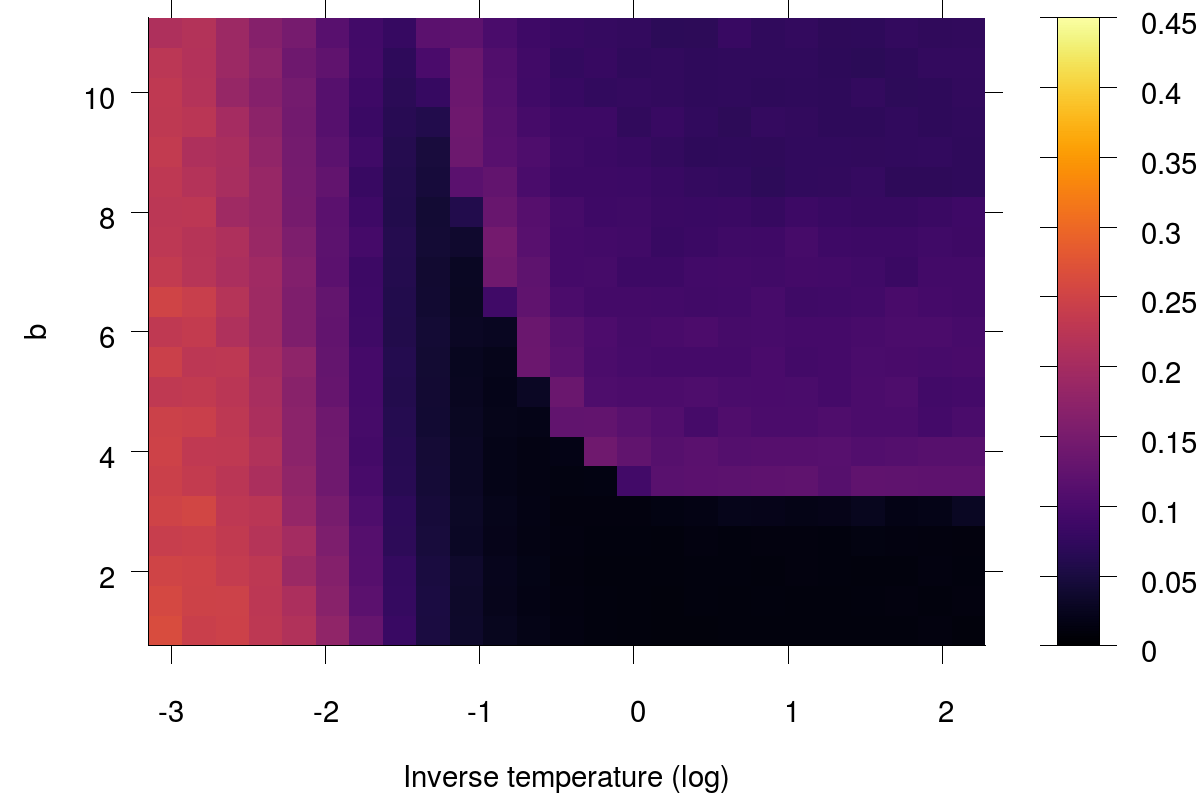}
  \caption{Inherited tags, strategy $(0,1)$}\label{fig:btempin01}
\end{subfigure}%
\begin{subfigure}{.45\textwidth}
  \centering
  \includegraphics[width=.9\linewidth]{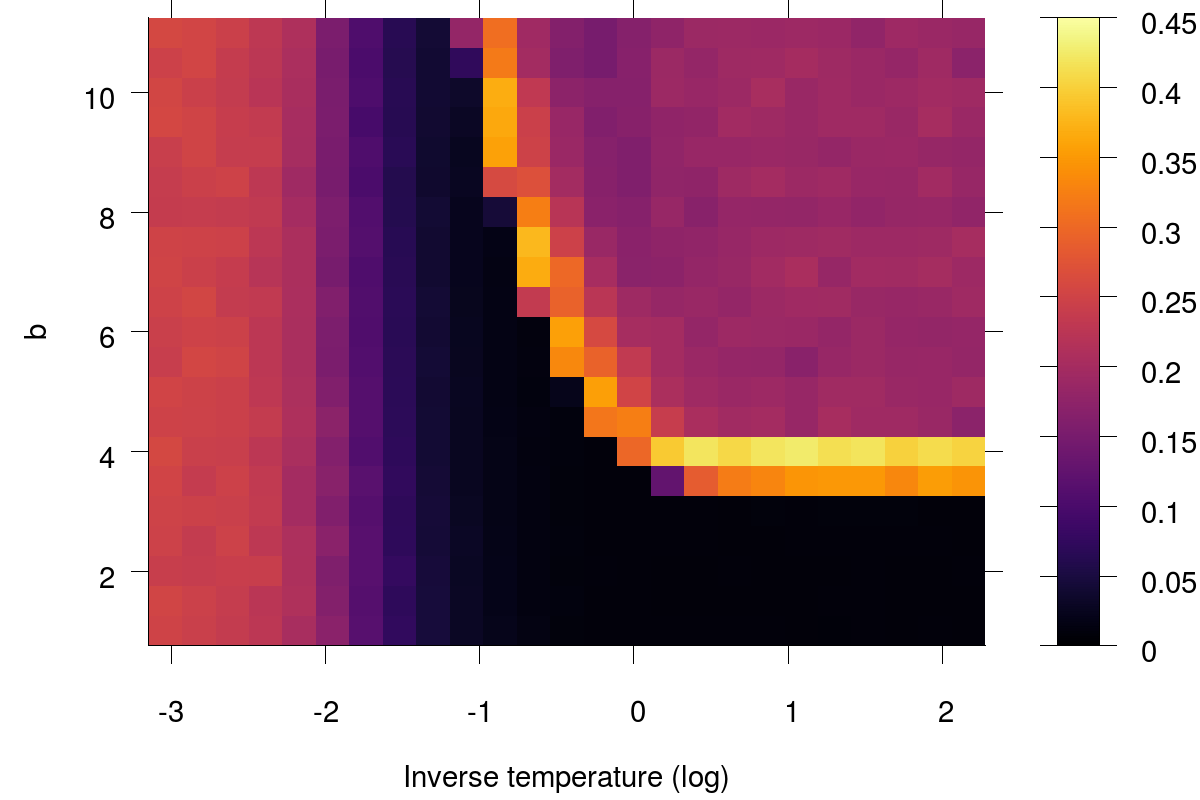}
  \caption{Random tags, strategy $(0,1)$}\label{fig:btempra01}
\end{subfigure}%

\begin{subfigure}{.45\textwidth}
  \centering
  \includegraphics[width=.9\linewidth]{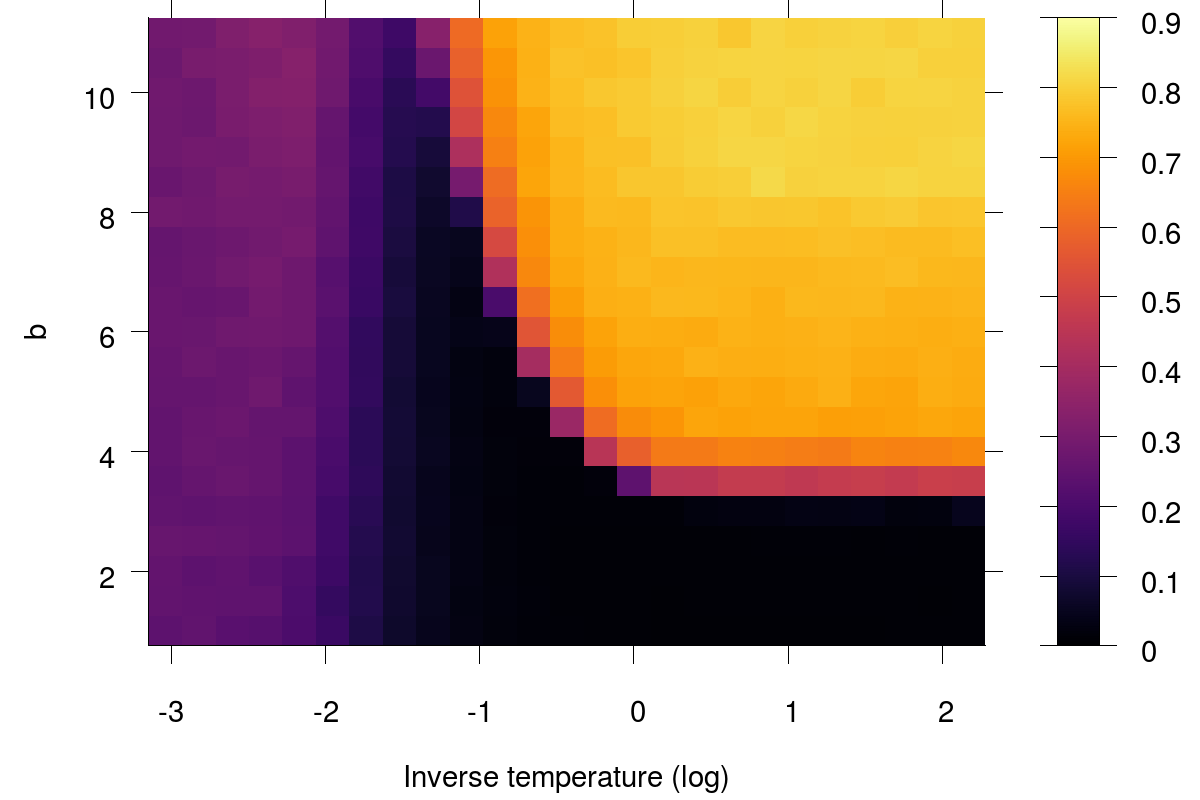}
  \caption{Inherited tags, strategy $(1,0)$}\label{fig:btempin10}
\end{subfigure}%
\begin{subfigure}{.45\textwidth}
  \centering
  \includegraphics[width=.9\linewidth]{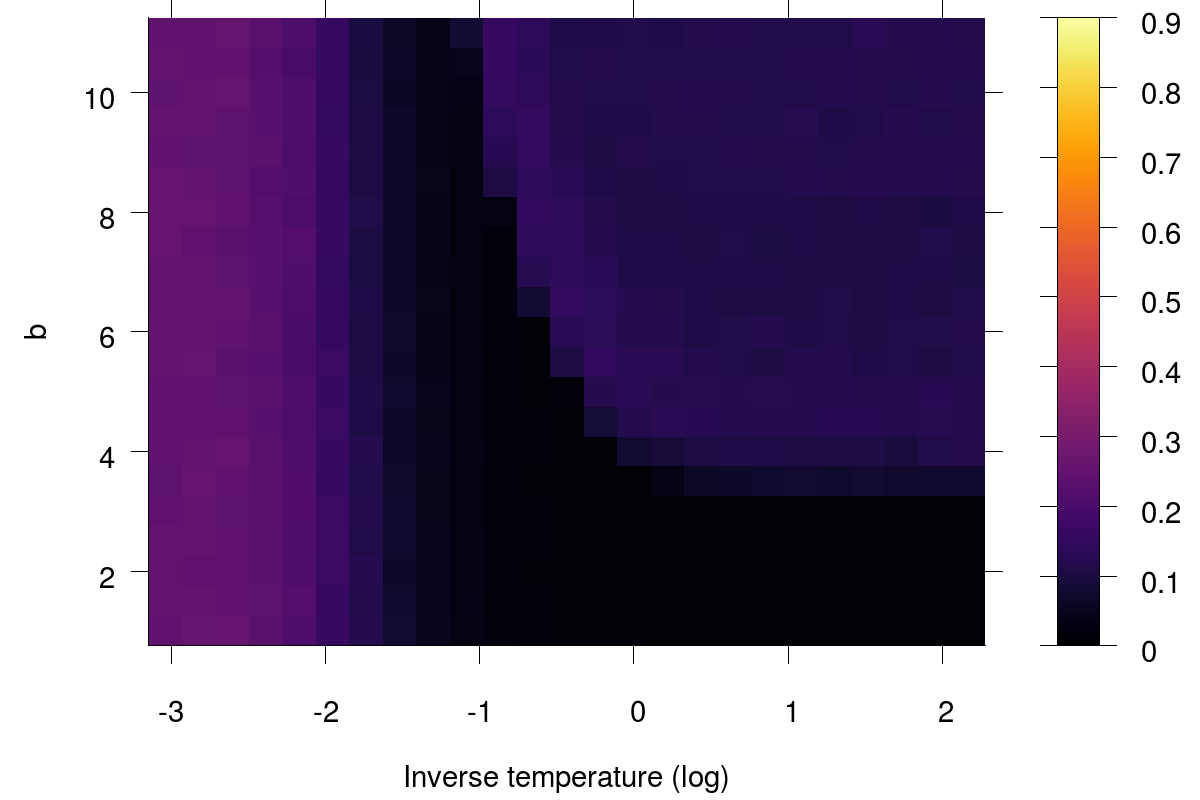}
  \caption{Random tags, strategy $(1,0)$}\label{fig:btempra10}
\end{subfigure}%

\begin{subfigure}{.45\textwidth}
  \centering
  \includegraphics[width=.9\linewidth]{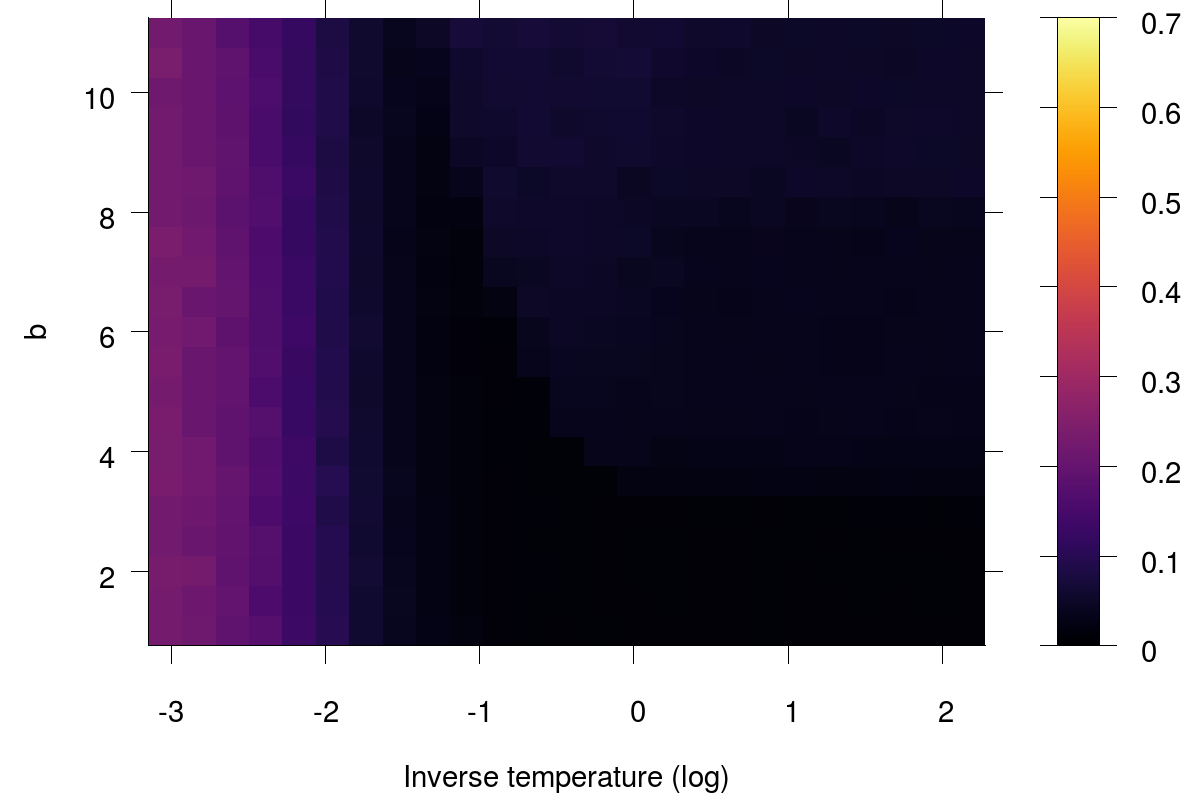}
  \caption{Inherited tags, strategy $(1,1)$}\label{fig:btempin11}
\end{subfigure}%
\begin{subfigure}{.45\textwidth}
  \centering
  \includegraphics[width=.9\linewidth]{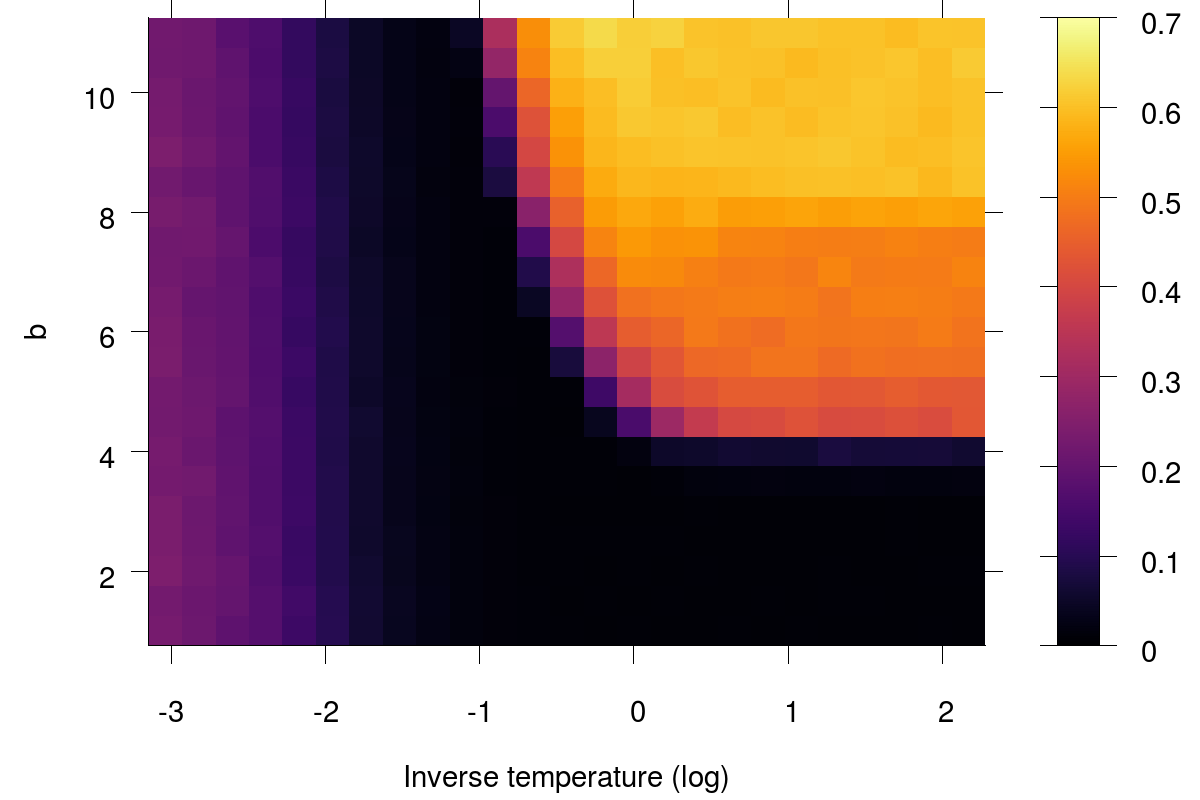}
  \caption{Random tags, strategy $(1,1)$}\label{fig:btempra11}
\end{subfigure}%

\caption{Proportion of each strategy (color scale) at generation 20,000 depending on $\beta$ and $b$ for the different treatments. Average proportions over 24 simulation runs.}\label{fig:btemp}
\end{figure}

\clearpage
\subsection{Change in strategy mutation probability}

%

\begin{figure}[ht]
\centering
\begin{subfigure}{.45\textwidth}
  \centering
  \includegraphics[width=.9\linewidth]{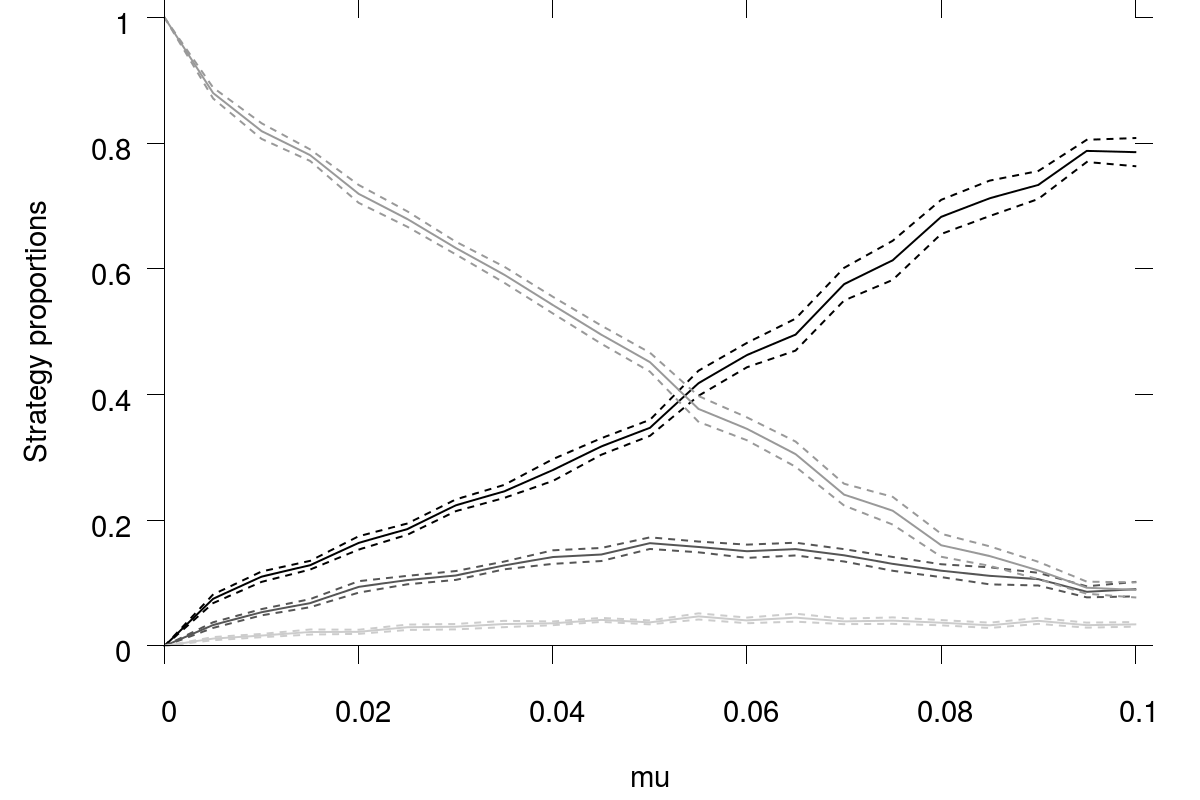}
  \caption{Inherited tags}\label{fig:6i2}
\end{subfigure}%
\begin{subfigure}{.45\textwidth}
  \centering
  \includegraphics[width=.9\linewidth]{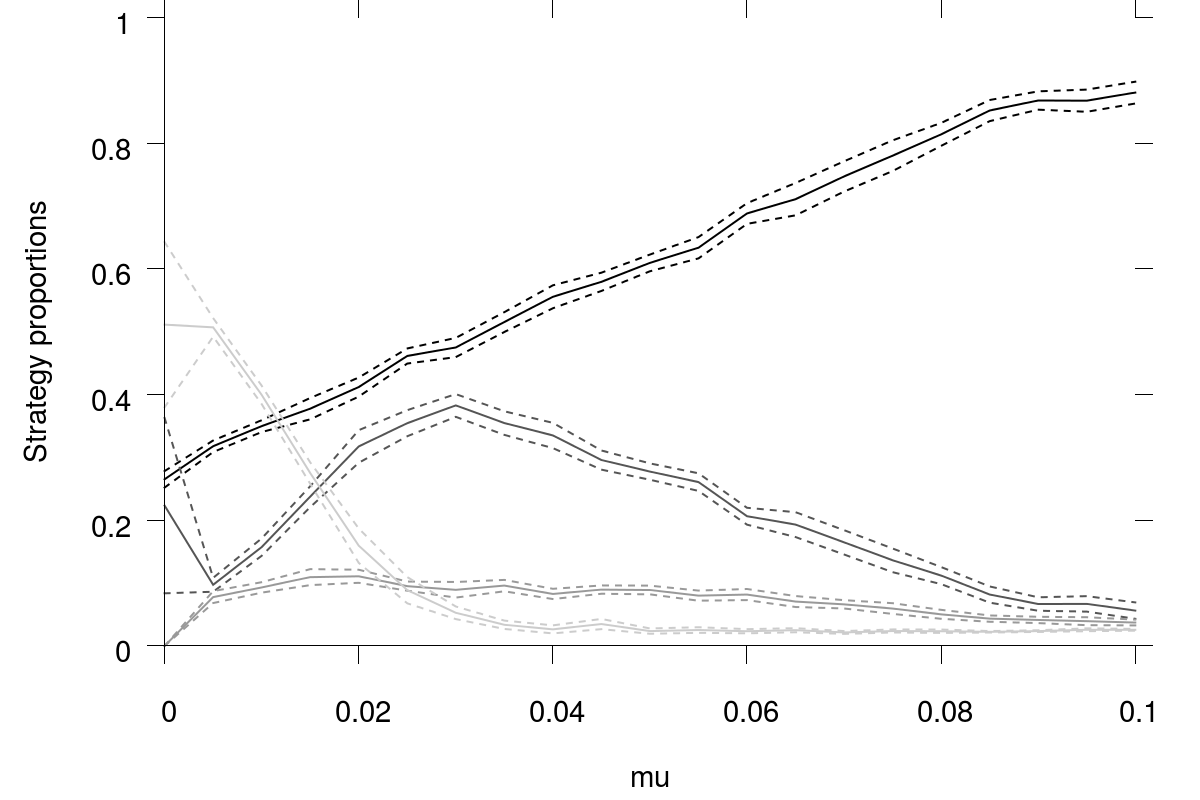}
  \caption{Random tags}\label{fig:6i1}
\end{subfigure}%
\caption{Evolution of the proportion of each strategy at generation 20,000 depending on $\mu$ and for the different treatments. From darkest to lightest grey: strategies $(0,0)$, $(0,1)$, $(1,0)$, $(1,1)$. Plain lines are averages and dashed lines are 95~\% confidence intervals for 24 simulation runs.}\label{fig:muvar}
\end{figure}

\begin{figure}[ht]
\centering
\begin{subfigure}{.45\textwidth}
  \centering
  \includegraphics[width=.9\linewidth]{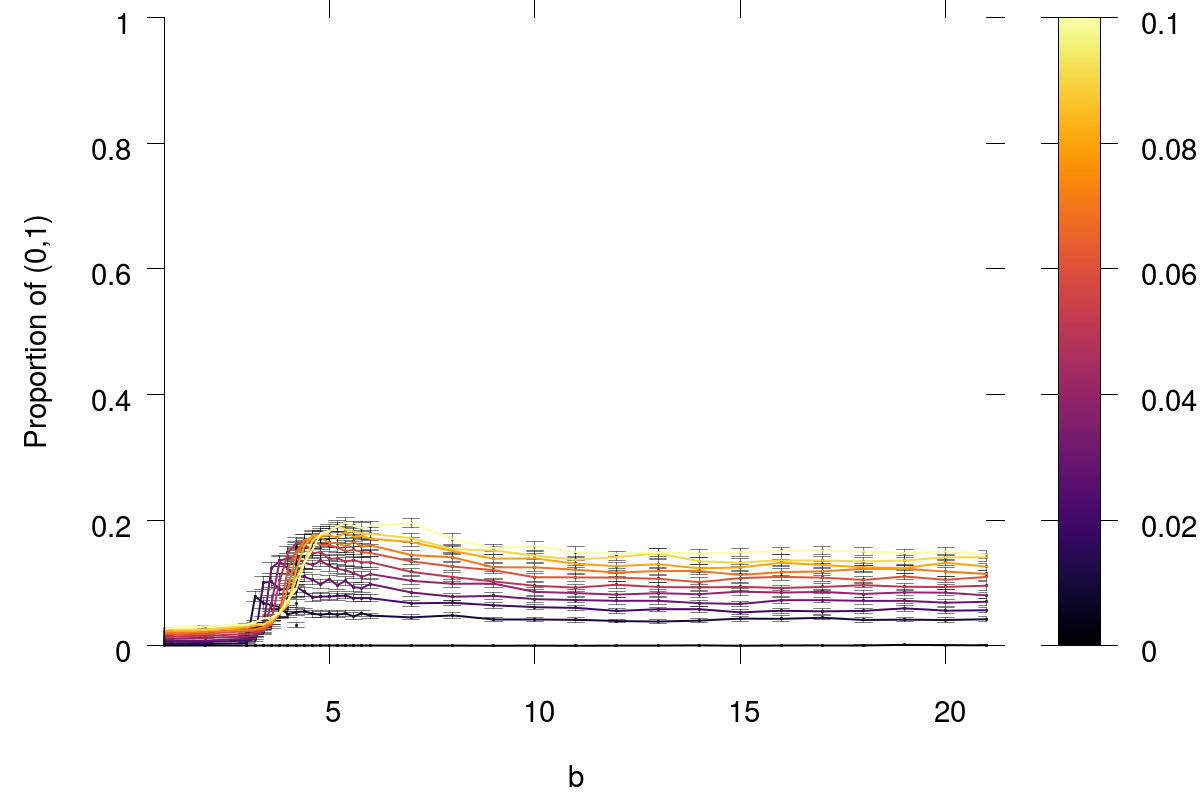}
  \caption{Inherited tags}\label{fig:5i2}
\end{subfigure}%
\begin{subfigure}{.45\textwidth}
  \centering
  \includegraphics[width=.9\linewidth]{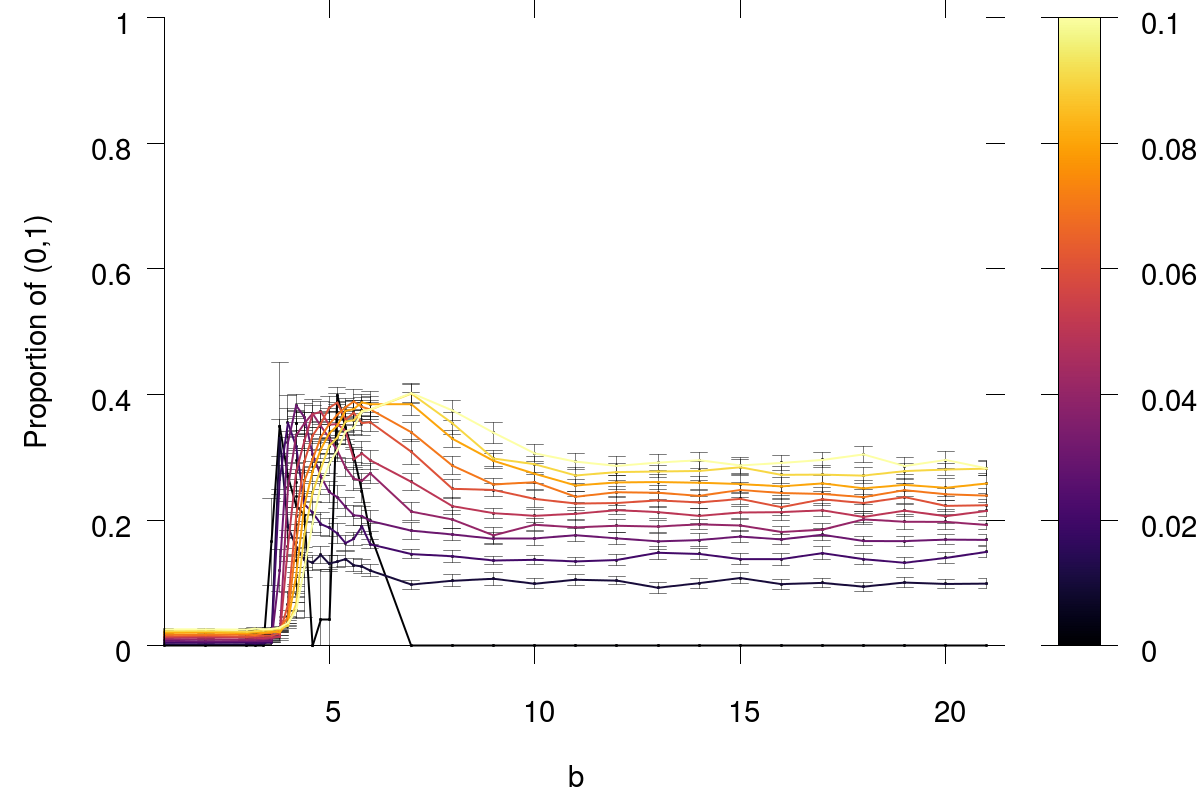}
  \caption{Random tags}\label{fig:5i1}
\end{subfigure}%
\caption{Proportion of players with strategy $(0,1)$ at generation 20,000 depending on $b$ and $\mu$ (color scale) and for the different treatments. Plain lines are averages and candle sticks are 95~\% confidence intervals for 24 simulation runs.}\label{fig:bmu}
\end{figure}

\clearpage
\subsection{Payoff correlations in the $2$-neighborhood}

\begin{figure}[ht]
\centering
  \includegraphics[width=.9\linewidth]{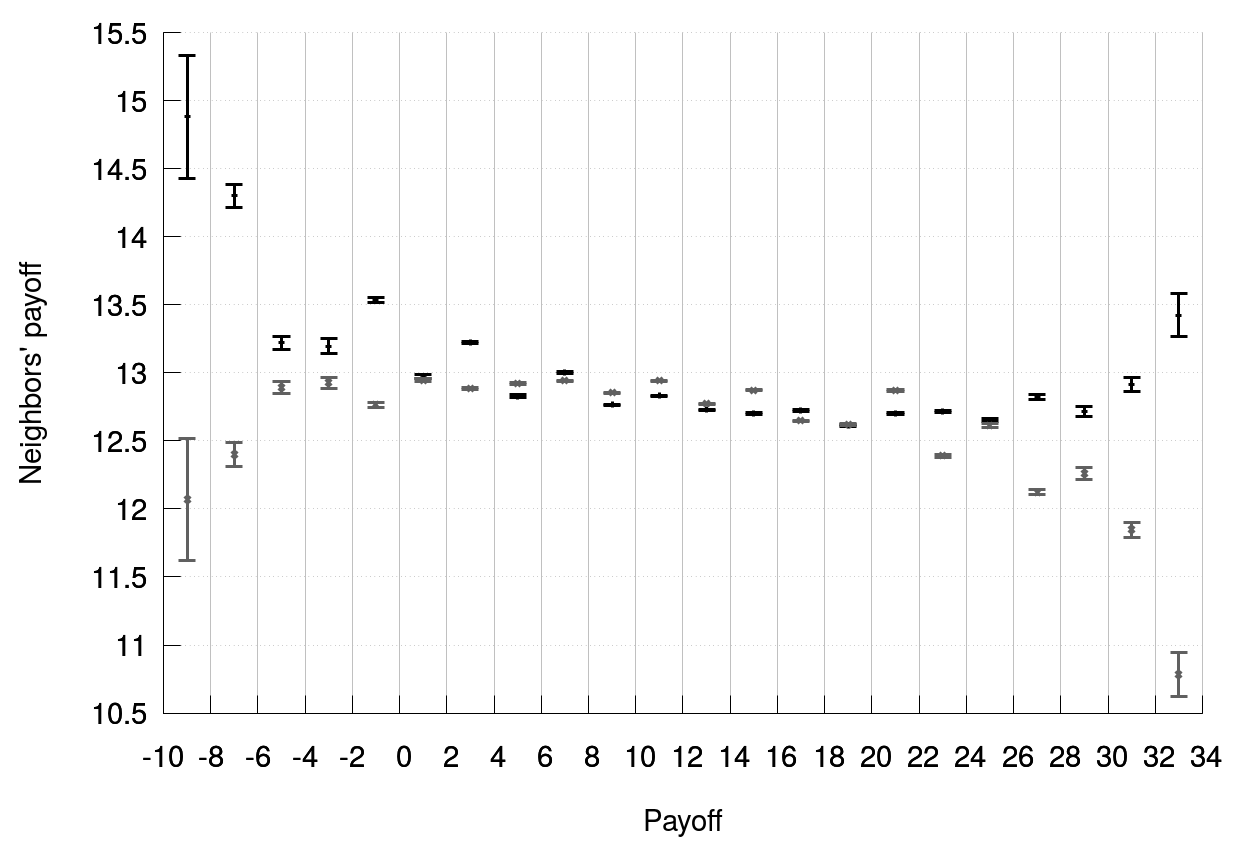}
\caption{Average payoffs in an agent's $2$-neighbors (excluding itself) as a function of the payoff of the agent itself, depending on it having strategy $(0,1)$ (black) or $(1,0)$ (grey). Whiskers: 95~\% CI with 1,000 simulations of the initial state. Vertical grey lines separate the payoff bins.}\label{fig:correl12}
\end{figure}

\end{document}

%% file: figs/fig1.tex
\begin{tikzpicture}

\draw[fill=black!70!white] (0.5cm,0.5cm) rectangle (3cm,3cm);

\draw[fill=black!20!white] (1cm,1cm) rectangle (2.5cm,2.5cm);

\draw[fill=black] (1.5cm,1.5cm) rectangle (2cm,2cm);
\draw[step=0.5cm,color=black,step=5mm] (0cm,0cm) grid (3.5cm,3.5cm);
\end{tikzpicture}

%% file: figs/tab1.tex
\begin{tabularx}{0.6\textwidth}{ccm{0.3\textwidth}}
\hline
\textbf{Parameter} & \textbf{Value} &\textbf{Description} \\
\hline
	$n$ & 50 & Grid width and length\\
	$\mu$ & 0.03 & Mutation probability \\
	$r$ & 1 & Neighborhoods size \\
	$\beta$ & 1 & Inverse temperature \\
	$c$ & 1 & Cooperation cost \\
	$b$ & $4.2$ & Cooperation benefit \\
	$T$ & 2 & Number of possible tags \\
	$G$ & 20,000 & Number of generations \\
\hline
\end{tabularx}